\documentclass[usenatbib]{mn2e}
\usepackage{graphicx,times}
\usepackage{bm}
\usepackage{epsf}

\def\rtf{r,\theta,\varphi}
\def\br{{\bf r}}

\def\n{\noindent}
\def\qtwo{\qquad\qquad}
\def\qthree{\qquad\qquad\qquad}

\def\myC{{\cal C}}

\def\myB{{\cal B}}

\def\ad{{_\delta}}
\def\ho{{\hat \Omega}}

\def\bk{{\bf k}}

\def\bl{{\bf l}}

\def\2p{{(2\pi)^2}}

\def\bl{{\bf l}}

\def\be{\begin{equation}}
\def\ee{\end{equation}}

\def\ben{\begin{eqnarray}}
\def\een{\end{eqnarray}}

\def\rpa{{r_{\parallel}}}
\def\rpe{{{\bf r}_{\perp}}}

\def\oh{{\hat\Omega}}
\def\nn{{\nonumber}}

\def\inte{\int_0^{\infty}}

\date{\today,~ $ $Revision: 0.9 $ $}

\begin{document}

\onecolumn

\title[Higher-order Statistics for 3D Weak Lensing]
{Higher-order Convergence Statistics for Three-dimensional Weak
Gravitational Lensing}

\author[Munshi, Heavens and Coles]
{Dipak Munshi$^{1,2}$, Alan Heavens$^{1}$ and Peter Coles$^2$ \\
$^{1}$Scottish Universities Physics Alliance (SUPA), Institute for Astronomy, University of Edinburgh, Blackford Hill,  Edinburgh EH9 3HJ, UK \\
$^{2}$School of Physics and Astronomy, Cardiff University, Queen's
Buildings, 5 The Parade, Cardiff, CF24 3AA, UK
\\}

\maketitle

\begin{abstract}
Weak gravitational lensing  on a cosmological scales can provide
strong constraints both on the nature of dark matter and the dark
energy equation of state. Most current weak lensing studies are
restricted to (two-dimensional) projections, but tomographic studies
with photometric redshifts have started, and future surveys offer the possibility
of probing the evolution of structure with redshift. In future we
will be able to probe the growth of structure in 3D and put tighter
constraints on cosmological models than can be achieved by the use of
galaxy redshift surveys alone. Earlier studies in this direction
focused mainly on evolution of the 3D power spectrum, but extension
to higher-order statistics can lift degeneracies as well as providing
information on primordial non-gaussianity. We present analytical
results for specific higher-order descriptors, the bispectrum and
trispectrum, as well as collapsed multi-point statistics derived
from them, i.e. {\em cumulant correlators}. We also compute quantities we call the
{\em power spectra associated with the bispectrum and trispectrum}, the
Fourier transforms of the well-known cumulant correlators. We
compute the redshift dependence of these objects and study
their performance in the presence of realistic noise and photometric
redshift errors.
\end{abstract}


\section{Introduction}
\label{sec:intro}

Until very recently, the best information about the power spectrum
of cosmological density perturbations has been obtained from
large-scale galaxy surveys and Cosmic Microwave Background (CMB)
observations. However, galaxy surveys only probe directly the
clustering of luminous matter, while CMB observations mainly explore
the power spectrum at a very early linear stage of their evolution.
Weak gravitational lensing studies  provide a complementary approach
for probing the cosmological power spectrum at modest redshift in an
unbiased way; for a recent review, see \cite{MuPhysRep08}. Weak
lensing is a relatively young subject; the first measurements were
published within the last decade
\citep{BRE00,Wittman00,KWL00,Waerbeke00}. Since then rapid progress
has been made on analytical modelling, technical specification and
the control of systematics. Over the course of the next few years,
photometric redshift surveys are going to be increasingly prevalent.
Such deep imaging surveys combined with resulting photometric
redshift information will mean that there will be a considerable
scope for using weak lensing studies to map the dark matter in the
universe in three dimensions.

Ongoing and future weak lensing surveys such as the CFHT
legacy survey{\footnote{http://www.cfht.hawaii.edu/Sciences/CFHTLS/}},
Pan-STARRS  {\footnote{http://pan-starrs.ifa.hawaii.edu/}}, the Dark
Energy Survey,  and further in the future, the Large Synoptic Survey
Telescope {\footnote{http://www.lsst.org/lsst\_home.shtml}}, JDEM
and Euclid will provide a wealth of information in terms of mapping
the distribution of mass and energy in the universe. With these
surveys, like the CMB, the study of
weak lensing is entering a golden age. However to achieve its full
scientific potential, control is needed of systematics such as arise
from shape measurement errors, photometric redshift errors, and intrinsic alignments.

The use of photometric redshifts to study weak lensing in three
dimensions was introduced by \citet{Heav03}. It was later developed
by many authors \citep{HRH00, HKT06, HKV07, Castro05}, and was shown
to be a vital tool in constraining dark energy equation of state
\citep{HKT06}, neutrino mass \citep{Kit08} and many other
possibilities. While the traditional approach deals with projected
surveys or with tomographic information, there has been substantial
recent progress in the development of studying weak lensing in 3D.


Early analytical work in weak lensing mainly adopted
a 2D approach  \citep{JSW00}, due to the lack of any
photometric redshift information about
the source galaxies. It is also interesting to note that most of
these studies also employed a flat-sky approach \citep{MuJai01,Mu00,MuJa00}, as the first generations
of weak lensing surveys mainly focused on small patches of the sky \citep{MuPhysRep08}.
These works made analytical predictions for lower-order moments as well
as the entire probability distribution function of convergence field $\kappa$ or shear $\gamma$
\citep{MuJai01,Valageas00, MuVa05,VaMuBa04,VaMuBa05}. A tomographic
(``2.5D'')approach has also been developed, wherein the sources are
divided into a few redshift slices
\citep{TakadaWhite03,TakadaJain04,Massey07,Schrabback09} and these slices are then
analyzed jointly, essentially using the 2D approach but keeping the
information regarding the correlation among these redshift slices. A
notable exception however in this trend was \citet{Stebbins96} who
developed the analysis techniques for weak lensing surveys covering
the entire sky. In recent years there has been a lot of interest in
developing analysis tools and predicting the cosmological impact of
future generations of weak lensing surveys with large sky coverage
which are naturally analyzed in the spherical harmonic domain. In
this paper, we present a very general study of 3D Weak lensing
beyond the power spectrum. Extending the formalism developed in
\citet{Heav03} and \citet{Castro05} we use higher-order statistics
to probe non-Gaussianity present in primordial anisotropy as well as
that induced by gravity.


While power spectrum analysis does provide the bulk of the
information regarding background cosmology, higher-order statistics
are useful to lift degeneracies,  allowing determination of
$\Omega_m$ and $\sigma_8$ independently - see, e.g.,
\cite{BerVanMell97,JainSeljak97, Hui99,Schneider98,TakadaJain03}.
Some of these studies also carried out tomographic analysis using
the bispectrum. Detailed Fisher matrix analysis found  that the
level of accuracy with which various cosmological parameters
including the dark energy equation of state parameters can be
enhanced considerably by using bispectrum data in combination with
power spectrum measurements. Higher-order studies also are important
for evaluating scatter in lower-order estimates; e.g. the
trispectrum is important for computing error bars in power spectrum
estimates \citep{TakadaJain09}. Most studies involving higher-order
correlations however mainly concentrate on projected convergence.
The aim of this paper is to extend higher-order analysis to 3D by
taking into account the radial distance as inferred through
photometric redshift information.


Modelling of the underlying mass distribution is necessary for
predictions of weak lensing multispectra. In earlier studies, the
hierarchical ansatz was found to be very useful in modelling
higher-order statistics of weak lensing observables
\citep{Fry84,Schaeffer84, BerSch92,SzaSza93, SzaSza97, MuBaMeSch99,
MuCoMe99a, MuCoMe99b, MuMeCo99, CMM99, MuCo00, MuCo02, MuCo03}.  The
hierarchical ansatz models the higher-order correlations as a
hierarchy, with higher-order correlation functions being expressed
as products of correlation functions. The diagrammatic
representation of these expressions resembles perturbative models of
the correlation hierarchy which typically develops with the onset of
gravitational clustering in collisionless media. The amplitudes of
these ``Feynman diagrams'' are of course different in the
perturbative regime and in the quasilinear regime. Various
hierarchical \emph{ansatze} differ in the way they assign amplitudes
to diagrams with different topologies. We will employ the most
generic hierarchical ansatz in modelling the underlying mass
distribution, and the method can be modified to take into account
any other specific forms of correlation hierarchy in a relatively
simple way.


Higher-order correlation functions have also been detected
observationally \citep{Hui99,BerVanMell97,BerVanMell02}. As expected
though these studies are more difficult than two-point as they can
be dominated by noise. There have been several studies in this
direction which focuses mainly on projected surveys as well as using
tomographic information \citep{Hu99,TakadaJain04,TakadaJain03,Semboloni08}.
Typically one-point cumulants or lower-order moments are employed to
compress information in higher-order correlation functions into a
single number. This is indeed due to the related gain in
signal-to-noise. A full analysis of multi-point correlation function
(or their respective Fourier transforms, the multispectra) is
relatively difficult because of the low signal-to-noise ratio of
individual modes. In their recent study \cite{MuHe09} used an
intermediate option: they  found that  a better approach, one that
is even optimal in certain cases, is to use the power spectra
associated with the various multi-spectra. These objects combine
various individual modes of the multi-spectra in a specific way and
can be computed from numerical simulations or observed data
relatively easily. While their work was motivated by cosmological
studies of non--Gaussianity involving the CMB, here we generalize it
to 3D weak lensing studies. We primarily focus here on convergence
studies but the results can be generalized to shear statistics. We
focus on three- and four-point statistics, but the formalism is
general. We develop the analysis tools and provide results both in
all-sky limit using a harmonics treatment (useful for future surveys
with large sky coverage) as well as using a patch-sky approach using
flat-sky Fourier transforms (for surveys with relatively small sky
coverage). In addition to full analytical results we also provide
results using the extended Limber approximation which can
drastically reduce the computational cost with very little loss of
accuracy at high wavenumbers.


The paper is arranged as follows: In \textsection2 we discuss the basic formalism
of 3D weak lensing and how it can be used to estimate the power spectrum in 3D.
It introduces the notations which will be used in the following sections. In \textsection3 we introduce the
models describing higher-order clustering and their associated hierarchical amplitudes which are then used to construct a model for
the bispectrum and higher-order multispectra in the nonlinear regime. In \textsection4
we focus on representation of bispectrum and trispectrum respectively in various coordinate systems.
In  \textsection5 we relate observed convergence statistics with the underlying statistics of
mass distribution. Sections \textsection6 and \textsection7 are devoted to development of the skew spectrum
 (3-point)  and the kurt-spectrum (4-point).   In \textsection7 we develop the formalism for surveys with small sky coverage and \textsection8 is reserved for discussion of results and future prospects. Throughout we will borrow the notations from \citet{Castro05} wherever possible.
We have analysed the effect of photometric redshift errors as an appendix.

The general formalism developed in this paper will have applicability in other areas of cosmology,
where 3D information can be used effectively, including future generations of 21cm surveys
as well as near all-sky redshift surveys.

\section{Weak lensing in 3D: Analysis of the 3D Power Spectrum}

In this section we introduce the formalism of weak lensing in 3D. It will also serve to
introduce the notations which we will use in the later sections. The formalism for 3D weak lensing
was developed by \citet{Heav03} and was developed further by several authors including \citet{Castro05}.
We will review results from \citet{Castro05} for power spectrum analysis here which will
be useful for introducing the notations and for the further developments presented in later sections.
We will also use the extended Limber approximation (to be introduced later; \cite{LoAf08}) to simplify results obtained in \citet{Castro05}.

In this work we will mainly be concerned with the convergence field. Analytical extensions to deal with shear fields
will be dealt with elsewhere. We introduce $\Phi(r,\theta,\varphi)$ as the 3D gravitational potential at a 3D position $\rtf$, and  $\phi(\br)$
the lensing potential. The radial distance $r(t)$ is related to the Hubble expansion parameter $H(t)= \dot a/a$
by $r(z) = c\int_0^{z}dz' / H(z')$. The Hubble parameter is sensitive to the contents of the Universe thereby
making weak lensing a useful probe to study dark energy. The line of sight integral relating the two potentials can be
written as \citep{Kaiser92}:

\begin{equation}
\phi(\br) \equiv \phi(r,\ho) = {2 \over c^2} \int_0^r dr' { f_K(r-r') \over f_K(r) f_K(r')} \Phi(r',\ho).
\end{equation}

\n
Derivation of the above expression assumes the Born approximation \citep{BerVanMell97,Schneider98,Waerbeke02}, which evaluates the line of sight integral along the unperturbed photon trajectory. Note that the lensing potential $\phi(r,\ho)$ is radially dependent.
Throughout these papers we denote vectors in bold letters, $c$ denotes the speed of light.
$r=r(t)$  is the comoving distance to the source at a given instance of time $t$  from the observer
who is situated at the origin ($r=0$). Depending on the background cosmology $f_K(r)$ can be
$\sin r, r,\sinh r$ for a closed $(K=1)$, flat $(K=0)$ or open $(K=-1)$ universes respectively.
Our convention for the Fourier transform for the 3D fields closely resemble that of \cite{Castro05}.
The eigenfunctions of the Laplacian operator in flat space when expressed in spherical coordinates
turn out to be a product of spherical Bessel functions $j_l(kr)$ in the radial direction and the spherical harmonics
on the surface of a unit sphere i.e. $Y_{lm}(\ho)=Y_{lm}(\theta,\phi)$. The eigenfunctions $j_l(kr)Y_{lm}(\theta,\phi)$
are associated with eigenvalues $-k^2$. The eigendecomposition and its inverse transformation can be expressed as:

\begin{equation}
\Phi_{lm}(k) = \sqrt{ 2 \over \pi}\int d^3r \Phi({\bf r})\, k \, j_l(kr) Y_{lm}^*(\ho); \qtwo
\Phi({\bf r}) = \sqrt{ 2 \over \pi} \int k dk \sum_{l=0}^{\infty} \sum_{m=-l}^{m=l}\Phi_{lm}(k)j_l(kr)Y_{lm}(\ho).
\end{equation}

\n
The choice for this eigendecomposition is  determined by various
factors.  First it can deal with large areal sky coverage, and secondly as the lensing is related to gravitational potentials, the expansion allows us easily to express the coefficients of expansion of the convergence (or shear) in terms of the expansion of the density field through the Poisson equation \cite{Heav03}.  $\Phi_{lm}(k)$ here is the spherical harmonic decomposition
of $\Phi({\bf r})$, and similarly for $\phi({\bf r})$. The orthogonality properties
for the harmonic modes for the 3D potential can be used to define the 3D all sky power spectra by the following expressions.

\begin{equation}
\langle \Phi_{lm}(k)\Phi_{l'm'}(k') \rangle =  C_l^{\Phi\Phi}(k)\delta_{1D}(k+k')\delta^K_{ll'}\delta^K_{mm'}; \qtwo
\langle \phi_{lm}(k)\phi_{l'm'}(k') \rangle =  C_l^{\phi\phi}(k)\delta_{1D}(k+k')\delta^K_{ll'}\delta^K_{mm'}.
\end{equation}

\n
Here the power spectrum $C_l^{\Phi\Phi}$ represents the 3D power spectrum associated with the 3D potential field $\Phi_{lm}(k)$ and
$\delta_{nD}$ and $\delta^K$  represent the n-dimensional Dirac and Kronecker delta functions respectively. The corresponding all-sky power spectra for the lensing potential $\phi$ is denoted by $C_l^{\phi\phi}$.  In comoving coordinates we can write:

\begin{equation}
\triangle \Phi(\br) = {3 \over 2 a} \Omega_m H_0^2 \delta(\br); \qtwo \Phi_{lm}(k;r) = -{3 \over 2 a(r) k^2} \Omega_m H_0^2 \delta_{lm}(k;r) =
{C \over a(r)k^2} \delta_{lm}(k;r).
\label{eq:kappa_den}
\end{equation}
\n Here $a(z)=1/(1+z)$ is the scale factor at redshift $z$,
$\Omega_m$ is the total matter density at $z=0$, and $H_0$ is the
Hubble constant today. $\delta_{lm}(k;r)$ is the eigendecomposition
of $\delta(\br)$.   When appearing after the semi-colon, the $r$ dependence (e.g. of $\Phi_{lm}(k;r)$) really
expresses the time-dependence of the potentials; see \cite{Castro05}
for a discussion of the subtleties of this. Any model we assume for
describing non-linear growth of perturbations for $\delta(\br)$ will
thus have direct impact on statistics of observed weak lensing
convergence $\kappa$ as it depends on $\delta(\br)$ through its
dependence on $\Phi(\br)$. The harmonic decomposition of the lensing
potential and the 3D gravitational potential are related by the
following expression \citep{Castro05}:

\be
\phi_{lm}(k) = {4 k \over \pi c^2 } \int_0^\infty dk' k' \int_0^{\infty} r dr j_l(kr) \int_0^r dr' \left [ {r-r' \over r'}\right ] j_l(k'r')\Phi_{lm}(k';r').
\label{phi2den}
\ee

\n
We ignore complications of varying radial selection function for the sources, and distance errors in the main text for clarity.   These are considered in Appendix A.

Using the relation which relates the density and the gravitational potential
as well as expressing the convergence field in terms of the lensing potential $\kappa_{lm}(k) = -{1 \over 2}l(l+1) \phi_{lm}(k)$;
we can express the convergence coefficients in terms of those describing the
density field. This is important because we can then relate the statistics of the density field with that of the
convergence field directly which is potentially observable:

\be
\kappa_{lm}(k) = {16 k A \over \pi c^2 }{1 \over l(l+1)} \int_0^\infty dk' k' \int_0^{\infty} r dr j_l(kr) \int_0^r dr' \left [ {r-r' \over r'}\right ] j_l(k'r')
{\delta_{lm}(k';r') \over k^2 a(r)}
\label{kappa2den}
\ee

\n
We have absorbed the cosmological constants in the constant $A = -{3 \Omega_M H_0^2/2} $
We will also introduce the quantity $I_l(k_i,k)$, which will be useful in displaying the future results:

\begin{equation}
I_l(k_i,k) \equiv k_i \int_0^{\infty} dr~r~j_l(k_i,r) \int_0^r dr'  \left ({r - r' \over r' } \right ) j_l(kr')
\sqrt{P^{\Phi\Phi}(k;r')}
\end{equation}

\n
We can now write down the power spectrum associated with the lensing potential $\phi$
in a more compact form. Next we relate the convergence power spectra in terms of the $C_l^{\phi\phi}$ using
their relationship in the harmonic domain.

\begin{equation}
C_l^{\phi\phi}(k_1,k_2)= {16 A^2 \over \pi^2 c^2}\int_0^{\infty} k^2 I_l(k_1, k)I_l(k_2,k)dk; \qquad
C_l^{\kappa\kappa}(k_1,k_2)= {1 \over 4} l^2(l+1)^2 C_l^{\phi\phi}(k_1,k_2).
\label{Cl_phi}
\end{equation}

\n
In deriving the above expression it was assumed that gravitational potential power spectrum can be
accurately approximated by $P^{\Phi\Phi} \sim \sqrt {P^{\Phi\Phi}(k;r) P^{\Phi\Phi}(k;r')}$. For a detailed
description and range of validity see \citet{Castro05}. Clearly the analysis outlined above follows three different steps.
First we relate the lensing potential $\phi_{lm}(k; r)$, or equivalently
$\kappa_{lm}(k; r)$, to the 3D potential $\Phi_{lm}(k;r)$.
Next the statistics of $\Phi_{lm}(k;r)$ are used to make
concrete predictions about the statistics of $\kappa_{lm}(k;r)$. However an intermediate step is required
to connect the 3D Fourier decomposition $\delta(\bk)$ and its harmonic counterpart $\delta_{lm}(k;r)$.  See \citep{Castro05} for details.

Throughout this paper, the analysis will rely on various assumptions.
For an arbitrary field $\Psi({\bf r};r)$ which is assumed isotropic and homogeneous with spherical harmonics
decomposition $\Psi_{lm}(k)$ can be characterized by a power spectrum $C_l(k;r)$.  It is important to realise that $r$, the comoving distance, also
plays the dual role of cosmic epoch, and we retain $r$ in harmonic representations to label the
cosmic epoch. The cross-power spectrum related to an arbitrary $3D$ field at two different radial distances (redshifts) $C_l(k;r,r')$ will
be expressed as  $\langle \Psi_{lm}(k;r)\Psi^*_{l'm'}(k';r') \rangle = C_l(k;r,r')\delta_{1D}(k-k')\delta^K_{ll'}\delta^K_{mm'}$.
It was shown by \citet{Castro05} that $C_l(k,r)$ is simply the 3D power spectrum $P(k;r)$, $C_l(k;r,r') = P(k;r,r')$.
For the derivation we need to expand the Fourier decomposition of $\Psi({\bf r};r)$ and exploit the fact
that $\langle\Psi({\bf k};r)\Psi({\bf k};r')\rangle = (2\pi)^3 P(k;r) \delta_{3D}({\bf k} - {\bf k}')$.
In our present analysis we will focus on extending these results to higher-order statistics. We model the
underlying statistics of the density field by using non-perturbative results and relate these
to the statistics of projected field such as convergence.
We also introduce power spectra related to multispectra to effectively compress the information content.
The results are presented both in harmonic space as well as in Fourier domain using the Fourier approximation.

\subsection{3D Convergence Power Spectrum Using the Limber Approximation}

Computations of higher-order multi-spectra are often difficult, given the multi-dimensional integrals involved, which
often make numerical computations time consuming if not prohibitive. The Limber approximation \citep{Limb54} or its generalization
to Fourier space is often used to simplify the evaluation numerically by reducing the dimensionality of the integrals.
Typically implementation of the Limber approximation is valid at  small angular separations which correspond to large
multipole moments $l$ in harmonic domain. It also requires smooth variations of the integrand compared to the Bessel functions of relevant $l$.
For a detailed description of various issues and calculations of next order correction terms see a recent discussion by \cite{LoAf08}.

We start by the following expression Eq.(\ref{phi2den}) and  Eq.(\ref{Cl_phi}) from the previous section:

\ben
C_l^{\phi\phi}(k_1,k_2) &=& {16A^2 \over \pi^2 c^2}\int k^2 dk I_l(k_1,k)I_l(k_2,k) \nn \\
 &=&  {16A^2 \over \pi^2 c^2} k_1 \int_0^\infty dr_a r_a j_l(k_1r_a) \int_0^r dr_a' \left [ {r_a -r_a' \over r_a'}\right ]
   k_2 \int_0^\infty dr_b r_b j_l(k_1r_b) \int_0^r dr_b' \left [ {r_b -r_b' \over r_b'}\right ] \nn \\
& &  \ \times \int k^2 dk \sqrt {P^{\Phi\Phi}(k;r_a')P^{\Phi\Phi}(k;r_b')} j_l(kr_a)j_l(kr_b)
\een

We now use the Limber approximation to simplify the $k$ integral which produces a $\delta_{1D}(r_a' -r_b')$
function. Integrating out $r_b'$ with the help of the delta function and renaming the dummy variable $r_a'$ to $r'$
we can finally write:

\ben
&& C_l^{\kappa\kappa}(k_1,k_2) = {\pi \over 2} {l^2(l+1)^2 A^2 \over 4}~k_1~k_2 \int_0^{\infty} dr_1 j_l(k_1r_1) \int_0^{\infty} dr_2 j_l(k_2r_2)I_l(r_1,r_2); \nn \\
&& \qtwo  I_l(r_1,r_2) = {16 \over \pi^2 c^2}\int_0^{r_{min}} ~dr'~ \left ( {r_1 - r' \over r'} \right )~ \left ( {r_2 - r' \over r'} \right )
\left( {l \over r' }\right)^2 P^{\Phi\Phi} \left ( {l \over r'};r' \right ); \qquad r_{min} = min(r_1,r_2).
\een

\n
The limits of the integral only cover the overlapping region. We notice here that if we use the Limber approximation
Eq.(\ref{eq:limber_approx2}), then this equation reduces to simpler form as
higher harmonics at different radial distances $r_i$ becomes uncorrelated.

\be
C_l^{\kappa\kappa}(k_1,k_2;r_1,r_2) = \delta_{1D}(r_1 - r_2)\left [ {\pi \over 2 r_1^2} \right ] {l^2(l+1)^2 \over 4}~k_1~k_2~I_l(r_1,r_2).
\ee

\n
The convergence power spectrum now can be computed using Eq.(\ref{kappa2den}). It is worth mentioning that this equation
establishes a direct link of convergence power spectra and the underlying mass distributions. In later sections we will
extend this result to higher-order multispectra.

The signal-to-noise associated with various estimators from all-sky weak lensing surveys and the issues related to optimization will be
dealt with in a separate work. In this paper we will focus mainly on development of statistics
which can be employed to study gravity-induced non-Gaussianity using higher-order statistics.
We will use the expressions for the $C_ls$ derived here for the construction of the
optimal estimators for the bispectrum and trispectrum in the following sections.

\section{Modelling Gravity-Induced Non-Gaussianity}

Two point statistics are useful to constrain cosmological parameters. However the convergence power spectrum
depends principally on a  specific combination of cosmological parameters. Typical additional inputs in the
form of external data sets such as the CMB, or higher-order statistics can lift the degeneracy.
While the use of the 3D power spectrum already tightens the constrains further gain is anticipated with the
help of non-Gaussianity studies in 3D.

It is already known from earlier studies \citep{TakadaJain04} that lensing tomography with the power spectrum and
bispectrum can act as a probe of dark energy and mass power spectrum. The lensing bispectrum
has a different dependence on the lensing weight function and the growth rate of perturbations. This is the main reason why bispectrum tomography can provide complementary constraints
to the power spectrum. In fact it was found in previous studies that constraints
from the bispectrum can be as tight as that from the power spectrum.

We will model the non-Gaussianity using the hierarchical ansatz
which is known to be a reasonable approximation at small scales in the highly nonlinear regime.
This approach has been used previously to model the statistics of the convergence and shear fields
\citep{MuJai01,Valageas00, MuVa05,VaMuBa04,VaMuBa05}.

\subsection{ The Hierarchical Ansatz in the Highly Non-linear Regime}

We need a reliable technique for modelling weak lensing statistics
beyond the power spectrum. On larger scales, where the density field
is only weakly nonlinear, perturbative treatments are known to be
valid. For a statistical description of dark matter clustering in
collapsed objects on small scales, the standard approach is to use
the halo model \citep{CooSeth02}.  However, an alternative approach
on small scales is to employ various \emph{ansatze} which trace
their origin to field theoretic techniques used to probe
gravitational clustering. For our work, we will use a hierarchical
\emph{ansatz} where the higher-order correlation functions are
constructed from the two-point correlation functions. Assuming a
tree model for the matter correlation hierarchy (typically used in
the highly non-linear regime) one can write the most general case,
the $N$ point correlation function, $\xi_N({\bf r}_1, \dots, {\bf
r}_n)$ as a product of two-point correlation functions
$\xi(|\br_i-\br_j|)$ \citep{Bernardreview02}. Equivalently in the
Fourier domain the multispectra can be written as products of the
matter power spectrum $P\ad(k_1)$. The temporal dependence is
implicit here.

\begin{equation}
\xi_N({\bf r_1}, \dots, {\bf r_n}) \equiv \langle \delta({\bf r_1})
\dots \delta({\bf r_n})\rangle_c = \sum_{\alpha,N=trees}
Q_{N,\alpha}\sum_{\rm labellings} \displaystyle \prod_{\rm edges
(i,j)}^{(N-1)} \xi(|\br_i-\br_j|).
\end{equation}

\n
It is however very interesting to note that a similar hierarchy develops in the quasi-linear regime at tree-level
in the limiting case of vanishing variance. However the hierarchical amplitudes become
shape-dependent in such a case. Nevertheless there are indications from numerical simulations that
these amplitudes become configuration-independent again as has been shown by high resolution studies
for the lowest order case $Q_3 = Q$ \citep{Scocci98, Bernardreview02}. In the Fourier space however
such an ansatz means that the entire hierarchy of the multi-spectra can be written in terms of
sums of products of power spectra with diferent amplitudes $Q_{N,\alpha}$ etc. , e.g. in the lowest order we can write:

\begin{eqnarray}
&& \langle \delta(\bk_1)\delta(\bk_2) \rangle_c  = (2\pi)^3 \delta_{3D}({\bf k}_1+{\bf k}_2)P(k_1)  \\
&& \langle \delta(\bk_1) \delta(\bk_2)\delta(\bk_3) \rangle_c = (2\pi)^3 \delta_{3D}(\bk_1+\bk_2+\bk_3)B(\bk_1,\bk_2,\bk_3) \\
&& \langle \delta(\bk_1) \cdots \delta(\bk_4) \rangle_c = (2\pi)^3 \delta_{3D}(\bk_1+\bk_2+\bk_3+\bk_4)T(\bk_1,\bk_2,\bk_3,\bk_4).
\end{eqnarray}

\n
The subscript $c$ here represents the connected part of the spectra. The Dirac delta functions $\delta_{3D}$
ensure the conservation of momentum at each vertex representing the multispectrum.

\begin{eqnarray}
&& B\ad(\bk_1,\bk_2, \bk_3)_{\sum \bk_i=0} = Q_3[P(k_1)P(k_2)+ P(k_1)P(k_3)+ P(k_2)P(k_3)]\\
&& T\ad(\bk_1,\bk_2,\bk_3,\bk_4)_{\sum \bk_i=0} = R_a[P(k_1)P(k_2)P(k_3) + cyc.perm.]+
R_b[P(k_1)P(|\bk_1+\bk_2|)P(|\bk_1+\bk_2+\bk_3|)+ cyc.perm.].
\end{eqnarray}

\noindent
Different hierarchical models differ in the way numerical values are allotted to various amplitudes. \citet{BerSch92}
considered ``snake'', ``hybrid'' and ``star'' diagrams with differing amplitudes at various order. A new ``star''  appears at each order.
higher-order ''snakes'' or ``hybrid'' diagrams are built from lower-order ``star'' diagrams. In models where we only have only star diagrams
\citep{VaMuBa04} the expressions for the trispectrum takes the following form:

\be
T(\bk_1,\bk_2,\bk_3,\bk_4)_{\sum \bk_i=0}  = Q_{4}[P(k_1)P(k_2)P(k_3) + cyc.perm.].
\ee
Following \cite{VaMuBa04} we will call these models ``stellar models''.
Indeed it is also possible to use perturbative calculations which are however valid only at large scales.
While we still do not have an exact description of the non-linear clustering of a self-gravitating medium
in a cosmological scenario, theses approaches do capture some of the salient features of gravitational
clustering in the highly non-linear regime and have been tested extensively aganinst numerical simulation in 2D
statistics of convergence of shear \citep{VaMuBa04}. These models were also used in modelling of the covariance of
lower-order cumulants \citep{MuVa05}.

\section{Theoretical Model ling of 3D Convergence Bispectrum for 3D Weak Lensing Surveys}

Previous studies of the bispectrum involving weak lensing include work on projection (2D) as well as tomography.
In most of these studies the prediction of the bispectrum is tied to a specific
assumption regarding the growth of instability in the underlying density distribution. The main
motivation for most of these studies was to put tighter constraints on the dark energy equation of state
using weak lensing surveys by lifting the degeneracy involved in power spectrum analysis alone.
The three-point correlation function in real space (or equivalently its harmonic transform the bispectrum) encodes
information about the departure from Gaussianity, and this departure can be induced by non-linear gravity or by non-Gaussian
initial conditions. We will focus on the gravity-induced bispectrum here and plan to present a complete treatment of the
effect of initial non-Gaussianity on 3D weak lensing statistics elsewhere.

\subsection{Linking the Density Bispectrum in Various Representations: $\delta(\bk)$, $\delta_{lm}(r)$ and $\delta_{lm}(k)$}

The convergence bispectrum in 3D will depend on modelling of underlying density bispectrum, We start by quoting the
relation of the 3D density bispectrum expressed in Cartesian coordinate and in the harmonic space,
we refer the reader to \cite{Castro05} for detailed derivation of the following equation:

\be
\delta_{lm}(k;r) = {1 \over \sqrt {2 \pi}}  k i^l \int d\Omega_k \delta(\bk; r) Y_{lm}(\Omega_k).
\ee

\n
Using this definition, we can link the bispectrum defined in Fourier space with the one in the harmonic domain:

\begin{eqnarray}
&& \langle \delta_{l_1m_1} (k_1;r_1)\delta_{l_2m_2}(k_2;r_2)\delta_{l_3m_3}(k_3;r_3)\rangle_c = \nonumber \\
&& \left ( {1 \over \sqrt{2\pi}}\right )^3  \{ k_1 k_2 k_3 \} i^{l_1+l_2+l_3} \int d\ho_{k_1} Y_{l_1m_1}(\ho_{k_1})
\int d\ho_{k_2} Y_{l_2m_2}(\ho_{k_2}) \int d\ho_{k_3} Y_{l_3m_3}(\ho_{k_3})
\langle  \delta({\bf k}_1;r_1) \delta({\bf k}_2;r_2) \delta({\bf k}_3;r_3) \rangle.
\end{eqnarray}

\n
Let us introduce the following notation for the bispectrum associated with the density field:

\begin{equation}
\langle \delta(\bk_1,r_1)\delta(\bk_2,r_2)\delta(\bk_3,r_3)\rangle = B(\bk_1, \bk_2, \bk_3;r_1,r_2,r_3) \delta_{3D}(\bk_1+\bk_2+\bk_3).
\end{equation}

\noindent
Expanding the Dirac delta function $\delta_{3D}(\bf {k_1 + k_2 + k_3} )$ and using Rayleigh's expansion of the exponentials:

\begin{eqnarray}
&& \delta_{3D}(\bk_1 + \bk_2 + \bk_3 ) =
{ 1 \over (2\pi)^3 } \int e^{i(\bk_1 +\bk_2 +\bk_3).{\bf r}} d^3 {\bf r}  \nonumber \\
&& = \sum_{L_i, M_i} \int d^3{\bf r} ~i^{L_1+L_2+L_3}~
 j_{L_1}(k_1r) j_{L_2}(k_2r) j_{L_3}(k_3r)
 Y_{L_1M_1}(\ho_{k_1})Y_{L_2M_2}(\ho_{k_2})Y_{L_3M_3}(\ho_{k_3})
 Y_{L_1M_1}(\ho) Y_{L_2M_2}(\ho) Y_{L_3M_3}(\ho).
\end{eqnarray}

\n
Next, we use the orthogonality property of the spherical harmonics, Eq.(\ref{ortho_spherical}) to carry out the integrals to simplify the
expression. We have introduced the notation $d^3\bk \equiv k^2\,dk d\ho_{\bf k}\equiv k^2\sin\theta_kdkd\theta_kd\phi_k$. This allows us to
write the bispectrum in a spherical harmonics representation to its Fourier counterpart. Spherical coordinates
are natural choice for various reasons. The line of sight integration can be treated quite separately with
sky coverage issues. As we will see the partial sky coverage issues can also be dealt with a natural way
in harmonic expansions. It is also important to notice that (radial) errors due to photometric redshift
can also be incorporated naturally  (see Appendix \ref{photo_z} for more details).

\begin{equation}
\langle \delta_{l_1m_1}(k_1;r_1))\dots\delta_{l_3m_3}(k_3;r_3)\rangle = \left ( {1 \over \sqrt{2\pi}}\right )^3  G_{l_1l_2l_3}^{m_1m_2m_3} \int r^2 d r j_{l_1}(k_1r)j_{l_2}(k_2r)j_{l_3}(k_3r) B(\bk_i;r_i).
\end{equation}

\n
The directional dependence through the azimuthal quantum number $m$ is encapsulated through the Gaunt integral $G$ and is
defined by the following expressions (we also introduce the quantity $I_{l_1l_2l_3}$ which we will find useful
later):

\ben
&& G_{l_1l_2l_3}^{m_1m_2m_3} = \int \,d\ho\, Y_{l_1m_1}(\ho) Y_{l_2m_2}(\ho)Y_{l_3m_3}(\ho) =
\left(
  \begin{array}{ c c c }
     l_1 & l_2 & l_3 \\
     m_1 & m_2 & m_3
  \end{array} \right)I_{l_1l_2l_3}; \\
&& I_{l_1l_2l_3} \equiv \sqrt{ (2l_1 + 1)(2l_2+1)(2l_3+1) \over 4 \pi  }  \left(
  \begin{array}{ c c c }
     l_1 &l_2 & l_3 \\
     0 & 0 & 0
  \end{array} \right).
\een

\begin{table*}
\caption{Notations in different bases for the bi- and trispectrum for the underlying mass distribution $\delta$. The respective
multispectra for the convergence field $\kappa$ will be denoted by corresponding calligraphic symbols ${\cal B}_{l_1l_2l_3}$ and
${\cal T}_{l_1l_2}^{l_3l_4}$.}
\label{tab_notation}
\begin{center}
\begin{tabular}{|c |c|c| c}
\hline
\hline
Function & 3pt Correlation  & 4pt Correlation & Basis/Space\\
\hline
\hline
$\delta(r)$ & $\xi_3$ ($r_1,r_2,r_3$) & $\xi_4(r_1,\dots,r_4)$&  Real Space\\
\hline
$\delta_{lm}(r)$ & $B^{\rm mixed}_{l_1l_2l_3}(r_1,r_2,r_3)$ & $T^{l_1l_2}_{l_3l_4}(L;r_i)^{\rm mixed}$ & $Y_{lm}$\\
\hline
$\delta_{lm}(k;r)$ & $B^{\rm sph}_{l_1l_2l_3}(k_1,k_2,k_3)$ & $T^{l_1l_2}_{l_3l_4}(L;k_i,r_i)^{\rm sph}$ & $Y_{lm}$; $j_l(kr)$ \\
\hline
$\delta(\bk;r)$ & $B^{\rm rect}(\bk_1,\bk_2,\bk_3)$ & $T^{\rm rect}(\bk_1,\bk_2,\bk_3,\bk_4)$ & $e^{i\bk\cdot{\bf x}}$ \\
\hline
\hline
\end{tabular}
\end{center}
\label{table:notation}
\end{table*}

\n
Here the matrices are the $3J$ symbols, which are nonzero only if the triplets of harmonics $(l_1,l_2,l_3)$
satisfy the triangle equality, including the condition that the sum $l_1+l_2+l_3$ is even
which ensures the parity invariance of the bispectrum. We will also need the shorthand
notation $I_{l_1l_2l_3}$  in our following derivations. The rotationally invariant bispectrum $B_{l_1l_2l_3}$ can
now be written  in terms of $B_{l_1l_2l_3}^{m_1m_2m_3}$ as:

\begin{equation}
B_{l_1l_2l_3}^{m_1m_2m_3}(k_i;r_i)^{\rm sph} =  \left(
  \begin{array}{ c c c }
     l_1 & l_2 & l_3 \\
     m_1 & m_2 & m_3
  \end{array} \right)
B_{l_1l_2l_3}^{\rm sph}(k_i;r_i); \qquad B_{l_1l_2l_3}(k_i;r_i)^{\rm
sph} = \sum_{m_1,m_2,m_3} \left(
  \begin{array}{ c c c }
     l_1 & l_2 & l_3 \\
     m_1 & m_2 & m_3
  \end{array} \right) B_{l_1l_2l_3}^{m_1m_2m_3}(k_i;r_i)^{\rm sph}.
\end{equation}

\n
We will also need the reduced bispectrum commonly used in the literature which has direct
correspondence to the flat-sky bispectrum.  In terms of the bispectrum $B_{l_1l_2l_3}$ we can define
$b_{l_1l_2l_3} = I_{l_1l_2l_3} B_{l_1l_2l_3}$. Finally we can write the general correspondence between
the spherical harmonics representation of the angular bispectrum
$B_{l_1l_2l_3}(k_i;r_i) = B_{l_1l_2l_3}(k_1,k_2,k_3;r_1,r_2,r_3),$ and its Fourier representation
$B_{l_1l_2l_3}(k_i;r_i)= B(k_1,k_2,k_3; r_1,r_2,r_3)$. We will suppress the explicit display of
the radial coordinates to simplify notations. We need to keep in mind in the Fourier representations the radial coordinates
simply denote the cosmic epoch.

\be B^{\rm sph}_{l_1l_2l_3}(k_i;r_i) = { \left ( 1 \over
\sqrt{2\pi}\right )}^{3} \left \{ k_1k_2k_3\right \}~
J_{l_1l_2l_3}(k_i;r_i) I_{l_1l_2l_3}; \qtwo b^{\rm
sph}_{l_1l_2l_3}(k_i;r_i) = { \left ( 1 \over \sqrt{2\pi}\right
)}^{3} \left \{ k_1k_2k_3 \right \}~ J_{l_1l_2l_3}(k_i;r_i).
\label{bispec_1} \ee

\n
The expression $J_{l_1l_2l_3}(k_i;r_i)$ encapsulates the dependence on $k_i$ and $r_i$ with $(i=1,2,3)$

\be J_{l_1l_2l_3}(k_i;r_i) = \int r^2dr B^{\rm rect}(k_1,k_2,k_3;
r_1,r_2,r_3)j_{l_1}(k_1r)j_{l_2}(k_2r)j_{l_3}(k_3r).
\label{eq:bispec_J} \ee

\n
In certain applications it is also interesting to work in a basis where the harmonic decomposition is carried out only on the surface of the
sky, retaining the radial dependence in configuration space. In such circumstances, the following transformations are useful in relating the bispectrum expressed in this {\em mixed} coordinate with
bispectrum in full spherical coordinate.

\be \delta_{lm}(r) = \sqrt {2 \over \pi} \int k dk j_l(kr)
\delta_{lm}(k); ~~ B_{l_1l_2l_3}^{\rm mixed}(r_1,r_2,r_3) = \left (
{2 \over \pi} \right )^{3/2} \int dk_1 k_1 j_{l_1}(k_1r_1) \dots
\int dk_3 k_3 j_{l_3}(k_3r_3) B_{l_1l_2l_3}^{\rm
sph}(k_1,k_2,k_3;r_1,r_2,r_3). \label{bispec_2} \ee

\n
The inverse transformation from the basis $\delta_{lm}(r)$ to $\delta_{lm}(k)$ and the related change in bispectrum are as follows:

\be \delta_{lm}(k) = \sqrt {2 \over \pi} \int r^2 dr k  j_l(kr)
\delta_{lm}(r); ~~ B^{\rm sph}_{l_1l_2l_3}(k_1,k_2,k_3;r_1,r_2,r_3)
= \left ( {2 \over \pi} \right )^{3/2} \int dr_1 r_1^2 k_1
j_{l_1}(k_1r) \dots \int dr_3 r_3^2 k_3 j_{l_3}(k_3r_3) B^{\rm
mixed}_{l_1l_2l_3}(r_1,r_2,r_3) \nonumber \label{bispec_3} \ee

\n
If we put the expression Eq.(\ref{bispec_1}) into the equation Eq.(\ref{bispec_2}) then we can finally write:

\be
B^{\rm mixed}_{l_1l_2l_3}(r_1,r_2,r_3) =  \left ( {2 \over \pi} \right )^{3/2} ~I_{l_1l_2l_3}
\int dk_1 k_1^2 j_{l_1}(k_1r_1) \dots
\int dk_3  k_3^2 j_{l_3}(k_3r_3) \int r^2 dr j_{l_1}(k_1r)\dots
j_{l_3}(k_3r) B^{\rm rect}(k_1,k_2,k_3;r_1,r_2,r_3).
\ee

\n
These relations, especially Eq.(\ref{bispec_1}) will be useful in linking the 3D bispectrum to the convergence bispectrum.
This is an extension of earlier results in \cite{Castro05} for the power spectrum, where
they showed $\myC_l(k;r_1,r_2)= P(k;r_1,r_2)$. The result at bispectrum level is more involved. In the next section we will
use some approximations to simplify these expressions.

\subsection{Linking the Convergence Bispectrum to the Underlying Matter Bispectrum}

To make contact with the observables we use the fact that the projected convergence (which is related to the lensing
potential) can be related directly to the 3D density field. We will start by linking the
3D convergence bispectrum ${\cal B}$ and the 3D density bispectrum expressed in harmonic coordinates. In the next
section we will express the bispectrum in spherical coordinate in terms of the bispectrum in rectangular coordinates
and use some well-motivated approximations to simplify the results. Using Eq.(\ref{phi2den}) we can write the following expression:

\ben && {\cal B}^{\rm sph}_{l_1l_2l_3}(k_i;r_i) = A^3 \left ( {4 k_1
\over \pi c^2 }\right ) \left ( {4 k_2 \over \pi c^2 }\right ) \left
( {4 k_3 \over \pi c^2 }\right )
\inte  { dk_1' \over k_1'} \inte dr_1 r_1 j_{l_2}(k_1'r_1') \int_0^{r_1} {dr_1' \over a(r_1')} \left [ {r_1-r_1' \over r_1'}\right ] \nn \\
&& \times \inte {dk_2' \over k_2'} \inte dr_2 r_2 j_{l_2}(k_2'r_2')
\int_0^{r_2} {dr_2' \over a(r_2')} \left [{r_2 - r_2' \over r_2'}
\right ] \inte {dk_3' \over k_3'} \inte dr_3 r_3 j_{l_2}(k_3'r_3')
\int_0^{r_3} {dr_3' \over a(r_3')} \left [ {r_3 - r_3' \over
r_3'}\right ] B^{\rm sph}_{l_1l_2l_3}(k_i';r_i'). \een

\n The  bispectrum ${\cal B}_{l_1l_2l_3}(k_i;r_i)^{\rm sph}$ is now
expressed in terms of the bispectrum ${B}^{\rm
sph}_{l_1l_2l_3}(k_i;r_i)$. This relation mixes modes only in radial
directions. On the surface of the sky there is no mixing of angular
harmonics corresponding to various $l$ values. While expressing the
density harmonics in terms of the 3D potential harmonics, we pick up
additional scale factor $a(r_i)$ and $k_i$ dependence in the
denominator (see Eq.(\ref{eq:kappa_den}) for more on notational
details).

We have so far ignored the presence of noise. Indeed because of the limited number of galaxies
available it may not be possible to probe individual modes of
the bispectrum at high signal-to-noise ratio. In later sections we will be able to address issues related to optimum combinations
of individual modes which may be better suited for observational studies.

\subsection{Specific Models for underlying bispectrum and Limber's Approximation to the Exact Analysis}

The above analysis relates the underlying 3D bispectrum to the 3D convergence bispectrum for the most general case. However numerical computation involving such multiple integrals can be prohibitive.
To make further progress we will use specific models of gravity-induced bispectrum to simplify
the calculations. We will also use the Limber approximation to simplify our results. The Limber
approximation is known to be a very good approximation for smaller angular scales or high $l$.
We would like to stress however that although the results presents here are for a specific
models for hierarchical clustering it is nevertheless possible to extend the results of our
analysis to other models too. Assuming the specific form of hierarchical ansatz,introduced before, we can have:

\begin{equation}
B^{\rm rect}(\bk_1,\bk_2,\bk_3; r_1,r_2,r_3) =
Q_3[P(k_1,r_1)P(k_2,r_2) + \rm{cyc.perm.}].
\end{equation}

\n
Using this notation for the function $J_{l_1l_2l_3}(r_1,r_2,r_3)$  we introduced in Eq.(\ref{eq:bispec_J}) takes the following form:

\begin{equation}
J_{l_1l_2l_3}(r_1,r_2,r_3) =  Q_3 ~ I_{l_1l_2l_3} \int r^2~dr ~j_{l_1}(k_1r)j_{l_2}(k_2r)j_{l_3}(k_3r)~\left [ P(k_1; r_1)P(k_2 ; r_2)+ {\rm cyc.perm.} \right ].
\end{equation}

\n
We use the extended Limber approximation (see Eq.(\ref{eq:limber_approx2})) to simplify the integrals involving $k'_i$.
The delta functions simplify the resulting $r'$ integrations,
Finally the observable convergence bispectrum can be written in terms of directly the density bispectrum as follows:

\ben && {\cal B}_{l_1l_2l_3}(k_i;r_i)^{\rm sph} =   \inte dr_1 r_1
j_{l_1}(k_1r_1) \inte dr_2 r_2 j_{l_2}(k_2r_2) \inte dr_3 r_3
j_{l_3}(k_3r_3)
{\cal I}_{l_1l_2l_3}(r_1,r_2,r_3), \nn \\
&& {\cal I}_{l_1l_2l_3}(r_1,r_2,r_3) \equiv\left [ {\pi \over 2} \right ]^3 Q_3 I_{l_1l_2l_3} \int_0^{r_{min}} dr \left [ {(r_1 -r ) \over a(r) r^3}\right ] \left [ {(r_2 -r ) \over a(r) r^3}\right ] \left [ {(r_3 -r ) \over a(r) r^3}\right ]
\left \{ P \left ({l_1\over r};r \right )P \left ({l_2\over r};r \right ) + \rm{cyc. perm.} \right \}.
\een

\n
The integral here extends to the overlapping region i.e. $r_{min} = min(r_1,r_2,r_3)$, and the final result is not specific
to the assumed non-Gaussianity, but assumes the hierarchical ansatz. For an arbitrary bispectrum the result can be expressed by a suitable change in
${\cal I}_{l_1l_2l_3}(r_1,r_2,r_3)$:

\be
{\cal I}_{l_1l_2l_3}(r_1,r_2,r_3) =\left [ {\pi \over 2} \right ]^3 I_{l_1l_2l_3} \int_0^{r_{min}} dr \left [ {(r_1 -r ) \over a(r) r^3}\right ] \left [ {(r_2 -r ) \over a(r) r^3}\right ] \left [ {(r_3 -r ) \over a(r) r^3}\right ]
B \left ({l_1\over r},{l_2\over r},{l_3\over r}; r,r,r \right )
\ee

For computation of the bispectrum in scenarios with a specific model for the primordial non-Gaussianity we
will have to replace the kernel that appears in the expression for  ${\cal I}_{l_1l_2l_3} (r_1,r_2,r_3)$ and
similar results will follow. In particular we can replace the gravity-induced bispectrum with models of the primordial bispectrum,  e.g.
$B^{loc}$ or $B^{equi}$, to compute the related bispectrum for convergence $\kappa$.

\section{Theoretical modelling of the Convergence Trispectrum for 3D Weak Lensing Surveys}

As before we start by linking the trispectrum in spherical
coordinates with the spatial trispectrum in rectangular coordinates.
The procedure we will follow will be very similar to what we have
done for the case of the bispectrum. We start by introducing the
trispectrum in the Cartesian coordinate $\langle \delta(k_1;r_1)
\dots \delta(k_4;r_4) \rangle = T^{\rm
rect}(k_i;r_i)\delta_{3D}(\bk_1+\bk_2+\bk_3+\bk_4)$ and in radial
and polar coordinates as:

\begin{equation}
\langle \delta_{l_1m_1}(k_1;r_1)\delta_{l_2m_2}(k_2;r_2)\delta_{l_3m_3}(k_3;r_3)\delta_{l_4m_4}(k_4;r_4)\rangle_c  = \sum_{LM} (-1)^M
  \left( \begin{array}{ c c c }
     l_1 & l_2 & L \\
     m_1 & m_2 & M
  \end{array} \right)
  \left(  \begin{array}{ c c c }
     l_3 & l_4 & L \\
     m_3 & m_4 & -M
  \end{array} \right) T^{l_1l_2}_{l_3l_4}(L,k_i;r_i)^{\rm sph}.
\label{eq:tri}
\end{equation}

\n
The vectors $l_1,l_2,l_3,l_4$ represents the sides of a quadrilateral and L is the length of the diagonal.
The matrices as before are the Wigner $3J$ symbols. The symbols are only non-zero when they satisfy
several conditions; which are $|l_1-l_2| \le L \le l_1+l_2$, $|l_3-l_4| \le L \le l_3+l_4$;
$l_1+l_2+L$= even, $l_3+l_4+L$ = even and $m_1 + m_2 = M$ as well as $m_3+m_4 = -M$.

In our notation for the trispectrum,
$T^{l_1l_2}_{l_3l_4}(k_i,r_i;L)$, the indices $(k_i,r_i) =
(k_1,r_1,\dots, k_4,r_4)$ encodes their dependence on various
Fourier modes of the density harmonics in the radial direction, used
in their construction. No summation will assumed over these variable
unless explicitly specified. We need also to subtract the Gaussian
or the disconnected part from the estimated trispectrum to compute
the connected part of the trispectrum, denoted by the subscript
$\langle \cdot \rangle_c$ in ensemble averaging. By expanding the
Dirac delta function in spherical harmonics and going through the
same algebra as above we can finally express
$T_{l_3l_4}^{l_1l_2}[L,k_i;r_i]^{\rm sph}$ in terms of
$T_4(k_i;r_i)^{\rm rect}$:

\be \langle
\delta_{l_1m_1}(k_1,r_1)\dots\delta_{l_4m_4}(k_4,r_4)\rangle_c =
\sum_{LM} (-1)^M G^{l_1l_2L}_{m_1m_2M} G^{l_3l_4L}_{m_3m_4-M} \int
r^2 dr j_{l_1}(k_1r_1)\dots
j_{l_4}(k_4r_4)T^{l_1l_2}_{l_3l_4}(L;k_i,r_i)^{\rm sph} \ee

\n
Next we express the four-point correlation function in terms of the trispectrum  $T_{l_3l_4}^{l_1l_2}[L,k_i;r_i]$. Finally
using the orthogonality properties of the 3J functions, we can finally connect the two representations. It involves
the functions $I_{l_1l_2l_3}$ we have introduced before. The prefactor involving $k_i$ is an artifact of the
normalization which we have adopted.

\be T_{l_3l_4}^{l_1l_2}[L;k_i,r_i]^{\rm sph} = \left ( {1\over
\sqrt{2\pi}} \right )^4~\left \{ k_1k_2k_3k_4 \right \}~\sum_L
I_{l_1l_2L}I_{l_3l_4L} J_4(k_i;r_i); \qquad J_4(k_i;r_i) \equiv \int
r^2 dr T_4(k_i;r_i)^{\rm rect} j_l(k_1r) \dots j_l(k_4r) \ee

\n
In our derivation we
have used the following identity to simplify the integration involving four spherical harmonics:

\be
\int d\ho Y_{l_1m_1}(\ho) Y_{l_2m_2}(\ho)  Y_{l_3m_3}(\ho)  Y_{l_4m_4}(\ho)=
\sum_{LM} (-1)^M G_{l_1l_2L}^{m_1m_2M} G_{l_3l_4L}^{m_3m_4-M}.
\ee

\n
We will also add the following expressions for the sake of completeness. As before we will relate the trispectrum defined from
the harmonics $\delta_{lm}(k;r)$ i.e. $T^{l_1l_2}_{l_3l_4}(L,k_i;r_i)$ with $T^{l_1l_2}_{l_3l_4}(L;r_i)$ which is defined from
the harmonics  $\delta_{lm}(k)$.

\be T^{l_1l_2}_{l_3l_4}(L;k_i,r_i)^{\rm sph} = \left ( { 2\over \pi
} \right )^2\int r_1^2  dr_1 k_1j_{l_1}(k_1r_1) \dots \int r_4^2
dr_4 k_4j_{l_4}(k_4r_4) T^{l_1l_2}_{l_3l_4}(L,r_i)^{\rm mixed} \ee

\n
The inverse relation which relates $T^{l_1l_2}_{l_3l_4}(L,r_i)$ with  $T^{l_1l_2}_{l_3l_4}(L;k_i,r_i)$ is given by following expression:

\be
 T^{l_1l_2}_{l_3l_4}(L,r_i)^{\rm mixed} =  \left ( { 2\over \pi } \right )^2\int k_1dk_1 j_l(k_1r) \dots \int k_4 dk_4 j_l(k_4r)T^{l_1l_2}_{l_3l_4}(L;k_i,r_i)^{\rm sph}
\ee

\n
The relation of the full 3D trispectrum in spherical coordinates and its Fourier decomposition, which generalizes our previous results for the bispectrum,
is written as follows:

\be
 T^{l_1l_2}_{l_3l_4}(L,r_i)^{\rm mixed} =
 \left ( {2 \over \pi} \right )^{2} \sum _L  I_{l_1l_2L} I_{l_3l_4L} \int dk_1 k_1^2 j_{l_1}(k_1r_1) \dots \int k_4^2 dk_4 j_{l_4}(k_4r_4)
 \int r^2 dr j_{l_1}(k_1r_1) \dots j_{l_4}(k_4r_4) T^{l_1l_2}_{l_3l_4}(L,k_i,r_i)^{\rm rect} .
\ee

\n Along with the bispectrum expression this generalizes the
previously-obtained relationship at the level of the power spectrum
in \citet{Castro05}. Clearly for practical purposes we will need to
devize an approximation to the exact result. We will use the Limber
approximation to approximate the spherical Bessel functions.  Note
that numerical evaluation of trispectra is considerably more
involved than the bispectrum. It is also important to note that as
we climb upwards in the hierarchy realistically it gets more
difficult to extract signals from observational data because of the
presence of noise.

\subsection{Linking the convergence trispectrum with the underlying matter trispectrum}

Finally the observable trispectrum ${\cal T}$ for the convergence $\kappa$ (defined through an equivalent expression as in Eq.(\ref{eq:tri})) can be expressed in terms of the underlying trispectrum of the mass distribution:

\ben && {\cal T}^{l_1l_2}_{l_3l_4}(k_i;r_i)^{\rm sph} = \left (
{4C  \over \pi c^2 }\right )^4 k_1k_2k_3k_4 \inte  { dk_1' \over k_1'} \inte dr_2 r_2 j_{l_2}(k_1'r_1') \int_0^{r_1} {dr_1' \over a(r_1')} \left [ {r_1-r_1' \over r_1'}\right ] \nn \\
&& \qquad \qquad \qquad \qquad \qquad \qquad   \dots \inte {dk_4'
\over k_4'} \inte dr_4 r_4 j_{l_4}(k_4'r_4') \int_0^{r_4} {dr_4'
\over a(r_4')} \left [{r_4 - r_4' \over r_4'} \right ]
T^{l_1l_2}_{l_3l_4}(k_i';r_i')^{\rm sph}. \een
The mode-mixing in spherical coordinates happens only in the radial direction. It is expected that the estimation
of the trispectrum from a realistic sky will be noise-dominated in the near future. This means estimation will be
difficult for individual modes. We will develop methods to compress the information content in individual modes in an optimal way
elsewhere. The trispectrum is dominated by the noise from galaxy intrinsic ellipticity as well as shot noise from the Poissonian nature of the galaxy
distribution.  To determine  this we need to take into fact that a contribution to the trispectrum not only comes
from the non-Gaussian signal but also from disconnected Gaussian terms too.

\subsection{Specific Forms for Underlying Matter Trispectrum and the Limber Approximation}

\n
We will derive the result quoted above for the case of the hierarchical ansatz with a ``stellar" approximation we make further use of the extended
Limber approximation to simplify. The hierarchical ansatz as well as the Limber approximation are both valid at the
small angular scale, which justifies their joint use to simplify the results. The result takes the following form:

\ben
&& {\cal T}^{l_1l_2}_{l_3l_4}(k_i;r_i) =   \inte dr_1 r_1 j_{l_1}(k_1r_1) \dots  \inte dr_4 r_4 j_{l_4}(k_4r_4)
{\cal I}_{l_1l_2l_3l_4}(r_1,r_2,r_3,r_4), \nn \\
&& {\cal I}_{l_1l_2l_3l_4}(r_1,r_2,r_3,r_4) \equiv \left [ {\pi \over 2} \right ]^3 \sum_L I_{l_1l_2L}I_{l_3l_4L} \int_0^{r_{min}} dr \left [ {(r_1 -r ) \over a(r) r^3}\right ] \dots \left [ {(r_4 -r ) \over a(r) r^3}\right ]
T^{l_1l_2}_{l_3l_4}(L;{l_i\over r};r_i).
\een

\n These results are extensions of analogous relations obtained for
the bispectrum. We will introduce contributions from star topology
under the stellar approximation (for other hierarchical ansatz\'e
see e.g. \citet{SzaSza93,SzaSza97} which assumes that the amplitudes
associated with all topologies are the same).

We are only concerned with the connected part of the trispectrum here.
Next we use the hierarchical ansatz to model the four-point correlation function.
We will only use the contribution from the diagram with ``star'' topology in this section.
The trispectrum expressed in Fourier domain, is simply the Fourier transformation of the four-point
correlation function. The trispectrum as outlined before in hierarchical approximation can be written as
a product of three power spectra:

\ben
\langle \delta(\bk_1,r_1)\delta(\bk_2,r_2)\delta(\bk_3,r_3)\delta(\bk_4,r_4)\rangle_c
&& = R_a \left [ \int d^3 \,\bk P(k_1)P(k_3)P(k) \delta_{3D}(\bf {k_1+k_2-k})\delta_{3D}(\bf {k_3+k_4+k} ) + \rm {cyc.perm.} \right ]  \nonumber \\
&& + R_b \left [ \int d^3 \,\bk  P(k_1)P(k_2)P(k_3) \delta_{3D}(\bf{(k_1+k_2-k})\delta_{3D}(\bf {k_3+k_4+k} ) + \rm{cyc.perm.} \right ].
\een

\n
In general the hierarchical amplitudes $R_a$ (associated with the snake topology) and $R_b$ (associated
with star topology) will have different amplitudes. There are $12$ terms with snake topology and $4$ terms with star topology
which are represented by the ``cyc.perm.''. Various hierarchical models differ in the way they ascribe values to various amplitudes.
It is possible also to employ Hyper Extended Perturbation Theory \citep{Scocci98} to compute these amplitudes. For our purpose we
will assume:

\be
\langle \delta(\bk_1;r_1)\delta(\bk_2;r_2)\delta(\bk_3;r_3)\delta(\bk_4;r_4)\rangle_c  = Q_4 \left [ \int \, d^3 \bk P(k_1)P(k_3)P(k) \delta_{3D}(\bf {k_1+k_2-k})\delta_{3D}(\bf {k_3+k_4+k} ) + \rm {cyc.perm.} \right ].  \nonumber
\ee
In this case, the analysis is essentially the same as that of the bispectrum and it simplifies the results considerably.
The stellar approximation consists of approximating the four-point correlation only with stellar diagrams. This model
has been checked in considerable detail in 2D in previous work \citep{BaMuVa04,MuVaBa04,VaMuBa05}. We will assume a ``stellar'' model from this point onward.
However the method outlined can also be generalised to take into account the ``snake'' diagrams. The cyclic permutations for the stellar model
now represents all $16$ diagrams.

We start by expanding the Dirac delta functions  $\delta_{3D}(\bk)$
using two dummy positional variables $x$ and $y$. Next following the same procedure as we have followed for the case of the bispectrum
we can express the star part of the
trispectrum in spherical coordinates. This will next be needed for the derivation of the convergence trispectrum.

\ben
\langle \delta_{l_1m_1}(k_1;r_1)\dots\delta_{l_4m_4}(k_4;r_4)\rangle_c \equiv
\int_0^{\infty} dr_1 r_1 j_{l_1}(k_1r_1) \dots \int_0^{\infty} dr_4 r_4 j_{l_4}(k_4r_4) J^{(4)}_{l_1l_2l_3l_4}(r_1,r_2,r_3,r_4),
\qthree \qthree  \nn \\
{J}^{(4)}_{l_1l_2l_3l_4}(r_1,r_2,r_3,r_4) \equiv Q_4 \left(\frac{4C}{\pi c^2}\right )^4 k_1 k_2 k_3 k_4
\sum_{LM} G_{l_1l_2L}^{m_1m_2M}G_{l_3l_4L}^{m_3m_4-M} \int r^2 dr j_{l_1}(k_1r) \dots j_{l_4}(k_4r)\left \{ P(k_1)P(k_2)P(k_3) + \rm {cyc.perm.} \right \},
\een

\n
where the definition for the kernel $J^{(4)}_{l_1l_2l_3l_4}(r_1,r_2,r_3,r_4)$ is similar to
its counterpart we introduced for the
bispectrum. In our notation, $A_{l_1l_2}^{l_3l_4}$
denotes the star contribution with the corresponding amplitude $R_a$. It is now possible to express the
star contribution to the 3D convergence trispectrum using the following relation:

\ben
 && \langle \kappa_{l_1m_1}(k_1;r_1) \dots \kappa_{l_4m_4}(k_4;r_4) \rangle_c =   \inte dr_1 r_1 j_{l_1}(k_1r_1) \dots  \inte dr_4 r_4 j_{l_4}(k_4r_4)
 {\cal I}_{l_1l_2l_3l_4}(r_1,r_2,r_3,r_4) \qthree \qthree \nn \\
 && {\cal I}_{l_1l_2l_3l_4}(r_1,r_2,r_3,r_4) = C^4 \left ( {4 k_1 \over \pi c^2 }\right ) \dots \left ( {4 k_4 \over \pi c^2 }\right ) \left [ {\pi \over 2} \right ]^3 \sum_L I_{l_1l_2L} I_{l_3l_4L} \int_0^{r_{min}} dr \left [ {(r_1 -r ) \over a(r) r^3}\right ] \dots \left [ {(r_4 -r ) \over a(r) r^3}\right ] \nn \\
&& \qquad \qquad \qquad \qquad \times \langle \delta_{l_1m_1}(k_1;r_1) \dots \delta_{l_4m_4}(k_4;r_4) \rangle_{star}.
\een

\n
We simplify the expression using the
Limber approximation as before. It effectively replaces the wavenumber $k_i$s with the corresponding $l_i/r$
while reducing the dimensionality of the integrals.

\ben
&& \langle \kappa_{l_1m_1}(k_1;r_1) \dots \kappa_{l_4m_4}(k_4;r_4) \rangle_c =
 \inte dr_1 r_1 j_{l_1}(k_1r_1) \inte dr_2 r_2 j_{l_2}(k_2r_2) \inte dr_3 r_3 j_{l_3}(k_3r_3)
{\cal I}_{l_1l_2l_3}(r_1,r_2,r_3) \nn \\
&& {\cal I}_{l_1l_2l_3}(r_1,r_2,r_3) =\left [ {\pi \over 2} \right ]^3 Q_4 \sum_L I_{l_1l_2L}I_{l_3l_4L} \int_0^{r_{min}} dr \left [ {(r_1 -r ) \over a(r) r^3}\right ] \dots \left [ {(r_4 -r ) \over a(r) r^3}\right ]
\left \{ P \left ({l_1\over r},r \right )P \left ({l_2\over r},r \right )P \left ({l_3\over r},r \right ) + \rm{cyc. perm.} \right \}.
\een

\n
Here the upper limit of integration along the radial direction  is $r_{min} = min(r_1,r_2,r_3,r_4)$.
In our analysis above we have taken the hierarchical ansatz as an example, but it is quite general and
the expression for the density trispectrum only affects the expression for $J^{(4)}_{l_1l_2l_3l_4}(r_1,r_2,r_3,r_4)$. Perturbative results
are in general more complicated to deal with because of configuration angle-dependence, but will result in  similar
signal-to-noise ratio. In stellar models the star topologies at
various orders carry all the weights, diagrams with snake topologies are ignored and arbitrary order in correlation functions are simply expressed in terms of the star contributions at that order. This approximation, as we will see, can simplify the calculations immensely.

We have concentrated here in modelling of the trispectrum and stressed its importance in confirmation of
detection of non-Gaussianity determined using the bispectrum. The analytical
modelling of the trispectrum is also important in itself for calculation of the error-covariance of the power spectrum.

\section{Convergence Skew-Spectrum}

\n Because of signal-to-noise issues it is difficult to study the
bispectrum for all possible configuration of triplets.  The skewness
compresses all the information contents of the bispectrum into a
single number. Such aggressive data compression may be elegant but
it fails to distinguish various contributions which might have
different shape dependence. These issues have been discussed
extensively in recent literature (see \citep{MuHe09} and references
therein for detailed discussion of related issues). The cumulant
correlators which were introduced in the literature are the
two-point objects and are well studied in the case of galaxy
surveys. These are muti-point statistics collapsed to two-point
objects. In harmonic space they correspond to power spectra
associated with multispectra of various orders. At the lowest order
there is only one power spectrum (coined the skew-spectrum) related
to the bispectrum. \citet{Cooray01} had earlier introduced the
unoptimised versions of power spectra associated with bispectra and
used them to study lensing-secondary cross correlation. Later
studies by \cite{Cooray06,CooLiMel08} used power spectra associated
with the bispectrum and trispectrum for redshifted 21cm studies.
These power spectra retain some of the information of the
multispectra that they are associated with, and the number increases
with the order. We will generalize and use the idea of cumulant
correlators here to study the bispectrum and trispectrum associated
with the 3D convergence field \footnote{Detailed modeling of a multispectra is
not important for defining the associated power spectra}.

The squared convergence field $\kappa^2(r_1)$ is constructed at a radial distance $r_1$; its
harmonic transform on the surface of the sky is denoted by $\kappa^{(2)}_{lm}(r_1)$ (we will be using the mixed representation throughout).
Let us start by expressing the spherical harmonics transform $\kappa_{lm}^{(2)}(r_1)$ of the squared
convergence field $\kappa^2(\hat\Omega,r_1)$ in terms of the
spherical harmonics of the original convergence map $\kappa_{lm}(r_1)$.

\be
\kappa^{(2)}_{lm}(r_1) = \int Y^*_{lm}(\hat\Omega)\kappa^2(\hat\Omega,r_1)d\hat\Omega = \sum_{l_1m_1} \sum_{l_2m_2} \kappa_{l_1m_1}(r_1) \kappa_{l_2m_2}(r_1) \int d \hat \Omega Y_{l_1m_1}(\hat \Omega)Y_{l_2m_2}(\hat \Omega)Y^*_{lm}(\hat\Omega) .
\ee

\n
The  above expression assumes all-sky coverage. In practice the surveys will cover only a fraction of the sky. The mask used in the survey
$w(\hat\Omega)$ will affect the estimators that introduces a multiplicative bias which needs to be properly accounted for. If we denote
the masked sky harmonics of the squared field by $\tilde\kappa^{(2)}_{lm}(r_1)$ then we can express them in terms of the original convergence harmonics of
the all-sky and the harmonics of the mask:

\ben
&&\tilde \kappa^{(2)}_{lm}(r_1) = \int w(\hat\Omega)Y_{lm}(\hat\Omega)\kappa^2(\hat\Omega,r_1)d\hat\Omega = \sum_{l_1m_1}\sum_{l_2m_2} \sum_{l_3m_3}
 \kappa_{l_1m_1}(r_1)\kappa_{l_2m_2}(r_1)  w_{l_3m_3} \int d \hat \Omega Y_{l_1m_1}(\hat \Omega)Y_{l_2m_2}(\hat \Omega)
Y_{l_3m_3}(\Omega) Y^*_{lm}(\Omega)  \\
&& \qtwo \equiv \sum_{(l'm')}K_{lml'm'}[w] \kappa^{(2)}_{l'm'}(r_1).
\een

\n Here the mixing matrix $K_{lml'm'}[w]$ denotes the mixing of
harmonics modes due to the presence of the sky mask whose
harmonic transform is $w_{lm}$. We will use the same mask at both
radial distances, but the results can easily be generalized to
two different masks.  Using these
harmonics we can now define the skew-spectrum (or the power spectrum
related to the bispectrum), $\myC_l^{(2,1)}$, as in \citet{MuHe09}.
The cut-sky version of the skew-spectrum is denoted by  $\tilde
\myC_l^{(2,1)}$ and is constructed from the cut-sky harmonics
$\tilde \kappa_{lm}^{(2)}$ and $\tilde \kappa_{lm}$ as follows:

\begin{equation}
C_l^{(2,1)}(r_1,r_2) \equiv {1 \over 2l+1} \sum_m \mathrm {Real} \left \{ \kappa_{lm}^{(2)*}(r_1) \kappa_{lm}(r_2) \right \} ; ~~~~~
\tilde C_l^{(2,1)}(r_1,r_2) \equiv {1 \over 2l+1} \sum_m \mathrm {Real} \left \{  \tilde \kappa_{lm}^{(2)*}(r_1) \tilde \kappa_{lm}(r_2) \right \}.
\end{equation}

\n
The skew-spectrum $\myC_l^{(2,1)}(r_1,r_2)$ probes directly the bispectrum  $B_{l_1l_2l_3}$. Though it does not
encode the entire shape dependence for each triangular configuration it encodes more information
compared to its one-point counterpart the skewness $S_3(r_1,r_2)$,  and has the ability to distinguish different
contributions to non-Gaussianity, both primordial as well as gravity-induced. We will focus on gravity-induced
non-Gaussianity here and issues related to primordial non-Gaussianity will be addressed elsewhere.

\begin{equation}
\myC_l^{(2,1)}(r_1,r_2) = \sum_{l_1,l_2} \myB_{ll_1l_2}(r_1,r_1,r_2) \sqrt{(2l_1+1)(2l_2+1)\over 4 \pi (2l +1) }
\left ( \begin{array}{ c c c }
     l_1 & l_2 & l_3 \\
     0 & 0 & 0
  \end{array} \right)
\label{eq:hat_c_l} = \frac{1}{2l+1} \sum_{l_1l_2} I_{ll_1l_2}\myB_{ll_1l_2}(r_1,r_1,r_2).
\end{equation}

\noindent
Here $B_{l_1l_2l_3}(r_1,r_2,r_3)$ is the angle-averaged bispectrum. It encodes
information about the three-point correlation function in the harmonic domain, $\langle \kappa_{l_1m_1}(r_1) \kappa_{l_2m_2}(r_1) \kappa_{l_3m_3}(r_2) \rangle_c$:

\begin{equation}
\langle \kappa_{l_1m_1}(r_1) \kappa_{l_2m_2}(r_2) \kappa_{l_3m_3}(r_3) \rangle_c = \myB_{l_1l_2l_2}(r_1,r_2,r_3)\left ( \begin{array}{ c c c }
     l_1 & l_2 & l_3 \\
     m_1 & m_2 & m_3
  \end{array} \right).
\end{equation}

\n
While the bispectrum is invariant under the permutations of the angular harmonics $(l_1,l_2,l_3)$
it is not invariant under the permutations of the radial distances $r_1,r_2,r_3$.
In certain circumstances the reduced bispectrum $b_{l_1l_2l_3}(r_1,r_2,r_3)$ is also useful to convey equivalent information:

\begin{equation}
B_{l_1l_2l_3}(r_1,r_2,r_3) = \sqrt {(2l_1+1)(2l_2+1)(2l_3+1)\over 4 \pi} \left ( \begin{array}{ c c c }
     l_1 & l_2 & l_3 \\
     0 & 0 & 0
  \end{array} \right)b_{l_1l_2l_3}(r_1,r_2,r_3) = I_{l_1l_2l_3}b_{l_1l_2l_3}(r_1,r_2,r_3).
\end{equation}

\n
For partial sky coverage one can obtain after tedious but straightforward algebra:

\ben
\tilde \myC_l^{(2,1)}(r_1,r_2) = && \sum_{l'}(2l'+1) \sum_{l''}{(2l''+1) \over 4 \pi}
\left ( \begin{array}{ c c c }
     l & l' & l'' \\
     0 & 0 & 0
  \end{array} \right)^2
|w_{l''}|^2  \sum_{l_1,l_2} \myB_{l'l_1l_2} (r_1,r_2,r_3) \sqrt{(2l_1+1)(2l_2+1) \over (2l'+1) 4 \pi}
\left ( \begin{array}{ c c c }
     l_1 & l_2 & l' \\
     0 & 0 & 0
  \end{array} \right)  \nn \\
&& = \frac{1}{2l+1}\sum I_{ll'l''}^2 |w_{l''}|^2 \left \{ \frac{1}{2l'+1} \sum_{l_1l_2} I_{l'l_1l_2} \myB_{l'l_1l_2}(r_1,r_1,r_2) \right \} \equiv M_{ll'} \myC_{l'}^{(2,1)}(r_1,r_2).
\een
\n
$w_{l}$ is the power spectrum of the mask, which is completely general. It is easy to see that in the  absence of any correlation between signal and
noise, the estimator $\myC_l^{(2,1)}$ is unbiased and no noise subtraction is needed as long
as it is Gaussian. Using the definition of the coupling matrix $M$, introduced above,
we express the pseudo-Cl (PCL) estimator as:

\begin{equation}
\hat \myC_l^{(2,1)}(r_1,r_2) = M^{-1}_{ll'} \tilde \myC_{l'}^{(2,1)}(r_1,r_2).
\end{equation}

\n
The covariance properties of such an estimator can be computed using similar techniques.
If we collapse the two-point correlator to a one-point object, we can write the cross-skewness as

\be
\hat S_3(r_1,r_2) = \sum_l (2l+1) \hat \myC_l^{(2,1)}(r_1,r_2).
\ee

\n
In our analysis we have so far only used the statistics $\langle \kappa^2(r_1)\kappa(r_2)\rangle$, which probes the bispectrum
$\myB_{l_1l_2l_3}(r_1,r_1,r_2)$. It is however also possible to consider the analogous statistics
$\langle \kappa(\oh,r_1)\kappa(\oh,r_2)\kappa(\oh',r_2)\rangle$ which probes
$\myB_{l_1l_2l_3}(r_1,r_2,r_2)$. There are  similar other permutations
which provide complementary information, with almost identical analysis.

\section{Kurt-Spectrum, or the power spectrum associated with the trispectrum}

The four-point correlation function, or its harmonic counterpart the
trispectrum, has been studied in the literature extensively. This
contains the information about the non-Gaussianity beyond the lowest
level \citep{Hu99,OkaHu02}. For the case of weak lensing studies
clearly the gravity-induced non-Gaussianity is the main motivation.
Studies in trispectrum analysis have also been pursued using various
other probes e.g. using 21cm surveys \citep{CooLiMel08} or more
extensively in several CMB studies; see \citet{BART04} for a review.
However these studies typically probe the trispectrum induced by
primordial non-Gaussianity.

It is important to note however that at the level of four-point studies, the
Gaussian fluctuations from the signal as well as the from the (Gaussian) noise too carry
a non-zero (unconnected) trispectrum. This degrades the signal-to-noise for various estimators
and clearly needs to be subtracted out before an unbiased comparison with the theoretical
predictions can be made.

It is obvious that detection of the trispectrum from noisy data is far more
nontrivial than the estimation of the bispectrum. Previous studies have mainly
concentrated on one-point estimators which collapse the data to a single
number - known as the kurtosis. We extend studies involving
kurtosis $\langle \kappa^4(\oh)\rangle$ to its two-point counterparts:
$\langle \kappa^2(\oh,r_1)\kappa^2(\oh',r_2)\rangle$ and
$\langle \kappa^3(\oh,r_1)\kappa(\oh',r_2)\rangle$. In practice however we will
consider the Fourier transforms of these objects which are the power spectra
 associated with the trispectra, $\myC_l^{(3,1)}$ and $\myC_l^{(2,2)}$. Indeed the radial
coordinates associated with two different fields being cross-correlated
can be different and will be denoted as $\myC_l^{(3,1)}(r_1,r_2)$ or
$\myC_l^{(2,2)}(r_1,r_2)$.

We start by defining the all-sky harmonic transform  $\kappa^{(3)}_{lm}(r_1)$ for the
convergence field $\kappa^3(\oh,r_1)$ and cross-correlate it against $\kappa_{lm}(r_2)$.
In the presence of a mask $w(\oh)$ which we assume to be the same at both radial distances,
the harmonic transforms of the cubic field $\tilde \kappa^{(3)}_{lm}(r_1)$ will depend also on the spherical
transforms of the mask $w_{lm}$ too:

\begin{eqnarray}
&& \kappa^{(3)}_{lm}(r_1) = \sum_{l_1m_1} \sum_{l_2m_2}\kappa_{l_1m_1}(r_1)\kappa_{l_2m_2}(r_1)\kappa_{l_3m_3}(r_1)
\int d \hat \Omega Y_{l_1m_1}(\hat \Omega)Y_{l_2m_2}(\hat \Omega)Y_{l_3m_3}(\hat \Omega)Y^*_{lm}(\hat \Omega) \\
&& \tilde \kappa^{(3)}_{lm}(r_1) = \sum_{l_1m_1} \sum_{l_2m_2} \sum_{l_3m_3}\kappa_{l_1m_1}(r_1)\kappa_{l_2m_2}(r_1)
\kappa_{l_3m_3}(r_1)w_{l_4m_4}
\int d \hat \Omega Y_{l_1m_1}(\hat \Omega)Y_{l_2m_2}(\hat \Omega)Y_{l_3m_3}(\hat \Omega)Y_{l_4m_4}(\hat \Omega) Y^*_{\
lm}(\hat \Omega)  \nn \\
&& = \sum_{l'm'}K_{lml'm'}\kappa^{(3)}_{l'm'}(r_1).
\end{eqnarray}
We will use these results to derive expressions for $\tilde \myC_l^{(3,1)}(r_1,r_2)$. The other
cut-sky power spectra $\tilde \myC_l^{(2,2)}(r_1,r_2)$ and all-sky counterparts $\myC_l^{(2,2)}(r_1,r_2)$ are
given by the following expressions:

\begin{equation}
\myC_l^{(2,2)}(r_1,r_2) = {1 \over 2l+1} \sum_m \kappa_{lm}^{(2)*} \kappa_{lm}^{(2)}  ~~~~~
\tilde \myC_l^{(2,2)} = {1 \over 2l+1} \sum_m  \tilde \kappa_{lm}^{(2)*} \tilde \kappa_{lm}^{(2)}.
\end{equation}

\n These power spectra directly probe
$T_{l_3l_4}^{l_1l_2}(l,r_1,r_2)^{\rm mixed}$. It compresses all the
available information in quadruplet of modes specified by
$(l_1,l_2,l_3,l_4)$ to a power spectrum. The power spectra
$C_l^{(2,2)}(r_1,r_2)$ and $C_l^{(3,1)}(r_1,r_2)$ differ in the way
they associate weights to various modes.

\ben \myC_l^{(2,2)}(r_1,r_2) = && \sum_{l_1,l_2,l_3,l_4}
T_{l_3l_4}^{l_1l_2}(l,r_1,r_2)^{\rm mixed} \sqrt
{(2l_1+1)(2l_2+1)\over 4\pi (2l+1)} \sqrt {(2l_3+1)(2l_4+1)\over
4\pi (2l+1)} \left (
\begin{array}{ c c c }
     l_1 & l_2 & l \\
     0 & 0 & 0
  \end{array} \right)
\left ( \begin{array}{ c c c }
     l_3 & l_4 & l \\
     0 & 0 & 0
  \end{array} \right) \nn \\
&& = \frac{1}{(2l+1)^2}\sum_{l_1l_2l_3l_4}  I_{lL_1l_2}I_{ll_3l_4}
T_{l_3l_4}^{l_1l_2}(l,r_1,r_2)^{\rm mixed} . \label{sphere22} \een

\n
Here the reduced trispectrum $T_{l_1l_2}^{l_3l_4}(r_i;L)^{mixed}$ is defined in terms of
$\langle \kappa_{l_1m_1}(r_1) \kappa_{l_2m_2}(r_2) \kappa_{l_3m_3}(r_3) \kappa_{l_4m_4}(r_4) \rangle_c$ as follows.
We have added the radial distances $r_i$ associated with each spherical harmonic
in the argument with $L$, which specifies the diagonal formed by the quadruplet of
four quantum numbers $l_i$.

\begin{equation}
\langle \kappa_{l_1m_1}(r_1) \kappa_{l_2m_2}(r_2)
\kappa_{l_3m_3}(r_3)\kappa_{l_4m_4}(r_4) \rangle_c = \sum_{LM}
T_{l_1l_2}^{l_3l_4}(r_i;L)^{\rm mixed} \left ( \begin{array}{ c c c
}
     l_1 & l_2 & L \\
     m_1 & m_2 & M
  \end{array} \right)
\left ( \begin{array}{ c c c }
     l_3 & l_4 & L \\
     m_3 & m_4 & -M
  \end{array} \right).
\end{equation}

\noindent
For partial sky coverage we can express the cut-sky version of the estimator  $\tilde C_l^{(2,2)}(r_1,r_2)$  in the
following way.  The resulting pseudo-$C_l$s can then be  expressed
in terms of $C_{l'}^{(2,2)}(r_1,r_2)$ through the mixing matrix $M_{ll'}$:

\begin{eqnarray}
\tilde \myC_l^{(2,2)}(r_1,r_2) &=& \sum_{l_1l_2l_3l_4} \sum_{l',l''} (2l'+1)
\left ( \begin{array}{ c c c }
     l & l' & l'' \\
     0 & 0 & 0
  \end{array} \right)^2
{ (2l'' +1 )\over 4\pi} |w_{l''}^2| \nonumber \\
&& \times ~ T_{l_3l_4}^{l_1l_2}(l,r_1,r_2)^{\rm mixed} \sqrt
{(2l_1+1)(2l_2+1)\over 4\pi (2l'+1)} \sqrt {(2l_3+1)(2l_4+1)\over
4\pi (2l'+1)} \left ( \begin{array}{ c c c }
     l_1 & l_2 & l' \\
     0 & 0 & 0
  \end{array} \right)
\left ( \begin{array}{ c c c }
     l_3 & l_4 & l' \\
     0 & 0 & 0
  \end{array} \right) \nn \\
&&  \qtwo = \frac{1}{2l+1}\sum_{l'l''}I^2_{ll'l''} |w_{l''}|^2 \left
\{ \frac{1}{2l+1}\sum
\frac{1}{2L+1}I_{l_1l_2L}I_{l_3l_4L}T_{l_3l_4}^{l_1l_2}(l,r_1,r_2)^{\rm
mixed} \right \}= \sum_{l'} M_{ll'}\myC_{l'}^{(2,2)}(r_1,r_2).
\label{partsphere22}
\end{eqnarray}

\noindent
There are two cumulant correlators at four-point level as explained above. Following the
discussion above we now focus on the  other degenerate power spectra associated with the
cumulant correlator $\kappa^3(\oh)\kappa(\oh')$. This is of the same order
as $\kappa^2(\oh)\kappa^2(\oh')$ and contains information about
trispectra as well. The compression of the information is done with different
weighting for different modes:

\begin{equation}
\myC_l^{(3,1)}(r_1,r_2) = {1 \over 2l+1} \sum_m \mathrm {Real} \left \{ \kappa_{lm}^{(3)*}(r_1) \kappa_{lm}^{(1)}(r_2) \right \} ~~~~~
\tilde \myC_l^{(3,1)}(r_1,r_2) = {1 \over 2l+1} \sum_m \mathrm {Real} \left \{\tilde \kappa_{lm}^{(3)*}(r_1) \tilde \kappa_{lm}^{(1)}(r_2) \right \}.
\end{equation}

\n
We can now use the definition of the trispectra $T_{l_1l_2}^{l_3l}(L;r_1,r_2)$ to express
$\myC_l^{(3,1)}(r_1,r_2)$ in terms of the trispectra. The main difference with the previous
spectrum $\myC_l^{(2,2)}(r_1,r_2)$ is that, it sums over all possible configuration
of the quadrilateral keeping one of the sides fixed, whereas $\myC_l^{(2,2)}(r_1,r_2)$
keeps one of the diagonal fixed but sums over all possible configuration of the
quadrilateral.

\ben \myC_l^{(3,1)}(r_1,r_2) && =  \sum_{l_1,l_2,l_3,L}
T^{l_1l_2}_{l_3l}(L,r_1,r_2)^{\rm mixed} \sqrt
{(2l_1+1)(2l_2+1)\over 4\pi \ (2L+1)} \sqrt {(2L+1)(2l_3+1)\over
4\pi (2l+1)} \left (
\begin{array}{ c c c }
     l_1 & l_2 & L \\
     0 & 0 & 0
  \end{array} \right)
\left ( \begin{array}{ c c c }
     L & l_3 & l \\
     0 & 0 & 0
  \end{array} \right) \\
&& =
\frac{1}{2l+1}\sum_{l_1l_2l_3;L}\frac{1}{2L+1}I_{l_1l_2L}I_{Ll_3l}
T^{l_1l_2}_{l_3l}(L,r_1,r_2)^{\rm mixed} . \label{sphere31} \een

\noindent
The partial sky coverage will mean that the measured power spectrum $\tilde \myC_l^{(3,1)}(r_1,r_2)$
is not the same as theoretical expectation, but is related as before by

\be
\tilde \myC_l^{(3,1)}(r_1,r_2)=M_{ll'}\myC_{l'}^{(3,1)}(r_1,r_2).
\label{partsphere31}
\ee

%

\n
In fact it can shown that for arbitrary sky coverage with arbitrary mask the above
analysis can be generalized to arbitrary order of correlation hierarchy. If we
consider a correlation function at $p+q$ order, for every possible combination of
$(p,q)$ we will have an associated power spectrum. Using the same expression for the
mode mixing matrix, we can invert the observed ${\tilde \myC}_{l}^{p,q}(r_1,r_2)$
to ${\hat \myC}_{l}^{p,q}(r_1,r_2)$.
Hence for arbitrary mask with arbitrary weighting functions
the deconvolved set of estimators can be written as:

\begin{eqnarray}
{\hat \myC}_l^{(p,q)}(r_1,r_2) = M_{ll'}^{-1} {\tilde \myC}_{l'}^{(p,q)}(r_1,r_2). \nonumber \\
\end{eqnarray}

\n
The Gaussian components of the corresponding multispectra at that order need
to be subtracted out, and can be written in terms of the $C_l$. The noise contribution too is
assumed Gaussian and hence only contributes to the unconnected components.
As before we can collapse the two-point objects and reduce them to a one-point number, the cross-kurtosis, which
will be a function of both radial distances $r_1,r_2$.

\be
K_4(r_1,r_2) = \sum_l (2l+1) \myC_l^{(3,1)}(r_1,r_2) = \sum_l (2l+1) \myC_l^{(2,2)}(r_1,r_2).
\ee
As we demonstrated with the cross-skewness, $K_4(r_1,r_2)$ can be decomposed in Fourier modes in the radial direction and
an associated power spectrum can be defined.

The Gaussian contribution to the trispectrum can be written as:

\ben
G^{l_1l_2}_{l_3l_4}(r_1,r_2,r_3,r_4;L) =
&& (-1)^{l_1+l_3} \sqrt {(2l_1+1)(2l_3+1)} \myC_{l_1}(r_1,r_2)\myC_{l_3}(r_3,r_4) \delta_{L0}\delta_{l_1l_2}\delta_{l_2l_3} \nonumber \\
&&  + (2L+1) (-1)^{l_2+l_3+L} \delta_{l_1l_3}\delta_{l_2l_4} \myC_{l_1}(r_1,r_3)\myC_{l_2}(l_2,r_4)
  +  (2L+1)\myC_{l_1}(r_1,r_4)\myC_{l_2}(r_2,r_3)  \delta_{l_1l_4}\delta_{l_2l_3}.
\een
Next we can compute the Gaussian contributions to $\myC_l^{(3,1)}$  and  $\myC_l^{(2,2)}$ following the same procedure
as before just by replacing the trispectrum $T^{l_1l_2}_{l_3l_4}$ with its Gaussian counterpart $G^{l_1l_2}_{l_3l_4}(r_i;L)$.
Indeed we will have to keep in mind the ordering correct for various $l_i$ and their $r_i$ counterparts.
It is also important to realize that in computing the Gaussian contribution we will have to take into account
both the signal and the noise $C_ls$ (assumed to be Gaussian), i.e, $C_l= C_l^S+C_l^N$.

\ben
&& G_l^{(3,1)}(r_1,r_2)
= \frac{1}{2l+1}\sum_{l_1l_2l_3;L}\frac{1}{2L+1}I_{l_1l_2L}I_{Ll_3l} G^{l_1l_2}_{l_3l}(L,r_1,r_2)^{\rm mixed} \nn \\
&& G_l^{(2,2)}(r_1,r_2) = \frac{1}{(2l+1)^2}\sum_{l_1l_2l_3l_4}
I_{ll_1l_2}I_{ll_3l_4} G_{l_3l_4}^{l_1l_2}(l,r_1,r_2)^{\rm mixed} .
\een

\n
For realistic surveys with a mask, results identical to Eq.(\ref{partsphere31}) and Eq.(\ref{partsphere22}) will hold true for the Gaussian contributions.
From the estimated $\tilde C_l^{(3,1)}(r_1,r_2)$ and $\tilde C_l^{(2,2)}(r_1,r_2)$  these contributions need to be subtracted out
before comparing them against the theoretical expectations.

\section{Optimal Estimates}

\n
The estimators introduced above are not optimal as they are not inverse-variance weighted. In this section
we discuss the signal-to-noise of the estimators introduced above, namely $\myC_l^{(2,1)}$ at the level
of bispectrum and $\myC_l^{(3,1)}$ and $\myC_l^{(2,2)}$ at the level of trispectrum.

For construction of the optimum estimates the harmonics $a_{lm}$
needs to be inverse covariance weighted \citep{SmZa06}. We refer to
\citet{MuHe09,Munshi_kurt} for a complete discussion which also the
requires presence of a linear term \citep{Crem06} in the case of absence
of spherical symmetry. The optimal estimates at the three-point level
in the all-sky limit and in the presence of constant variance noise can be expressed as:

\be
\hat {\cal S}_l^{(2,1)}(r_1,r_2) = {1 \over 2l+1}\sum_{l_1l_2} \hat \myB_{ll_1l_2}(r_1,r_1,r_2)\myB_{ll_1l_2}(r_1,r_1,r_2) \left \{ {\myC_{l}(r_1,r_1)}^{-1}
\myC_{l_1}(r_2,r_2)^{-1} \myC_{l_2}(r_3,r_3)^{-1} +\rm cyc.perm \right \}.
\ee

\n
We have denoted the estimators by $\hat{}$. The other terms can be obtained by circular permutation.
Different choices of ${r_1,r_2,r_3}$ can give us different skew spectra. In our discussion of the unoptimized
version we have developed one specific example which corresponds to
$S_l(r_1,r_1,r_2)$ and was denoted by $C_l^{(2,1)}(r_1,r_2)$. The other choices that we can construct
are $S_l(r_1,r_2,r_1)$ and $S_l(r_1,r_2,r_2)$. It is important to note that $C_l^{(2,1)}(r_1,r_2)$  is not
invariant under permutation of its indices.

Similar results hold for the case of trispectrum. The optimal estimates for the case of the trispectrum, we can write
for the full sky estimates with uniform noise:

\ben
&& \hat {\cal K}_l^{(3,1)}(r_1,r_2) = {1 \over 2l+1}\sum_{l_1l_2l_3;L} {1 \over 2L+1 } \hat T^{l_1l_2}_{l_3l}(L;r_i) T^{l_1l_2}_{l_3l}(L;r_i) \left \{
 {\myC_{l_1}(r_1,r_1)}^{-1}{\myC_{l_2}(r_1,r_1)^{-1}}{\myC_{l_3}(r_1,r_1)^{-1}}{\myC_{l}(r_2,r_2)}^{-1} +\rm cyc.perm. \right \} \\
&& \hat {\cal K}_l^{(2,2)}(r_1,r_2) = {1\over 2l+1}\sum_{l_1l_2l_3l_4} {1 \over 2l+1} \hat T^{l_1l_2}_{l_3l_4}(l;r_i)
T^{l_1l_2}_{l_3l_4}(l;r_i)\left \{ {\myC_{l_1}(r_1,r_1)}^{-1}
 {\myC_{l_2}(r_1,r_1)^{-1}}{\myC_{l_3}(r_2,r_2)^{-1}}{\myC_{l_4}(r_2,r_2)^{-1}} +\rm cyc.perm. \right \}.
\een

\n
The one-point estimator with inverse covariance weighting is simply the sum over the free index  of the respective two-point
estimators and is expressed as follows:

\be
\hat {\cal S}_3(r_1,r_2) = \sum_l (2 l+1)\hat {\cal S}_l^{(2,1)}(r_i); ~~~~ \hat {\cal K}_{4}(r_1,r_2) = \sum_l (2 l+1)\hat {\cal K}_l^{(3,1)}(r_i) = \sum_l(2 l+1)\hat {\cal K}_l^{(2,2)}(r_i) .
\ee

\n
Depending on various choices to identify the quadruplet of the radial distances $r_i$ that are associated with
each harmonic $l_i$ we will have a different estimator which can provide complementary information.
These estimators can help us to optimize survey depth and width for the study of non-Gaussianity at a given order.

\section{Flat Sky Treatment}

As pointed out before for surveys with large opening angle, the all sky expressions developed so far involve the
expansion in terms of the spherical harmonics and spherical Bessel functions. However
for surveys which cover only a small patch of the sky most of the signal
comes from higher harmonics. In such a situation, a natural choice would be to directly
deal with a two-dimensional Fourier expansion suitable for flat space. This makes the analysis more straightforward.
Our analysis here closely follows that of \cite{Hu99} and \cite{OkaHu02}. We are however required to take into account the additional radial
coordinate in our analysis. We expand the 3D convergence field as before both in the line-of-sight
direction as well as on the surface of the sky:

\be
\kappa(\bl,k) = \sqrt{2 \over \pi} \int d\rpa \int {d^2\rpe \over 2\pi} \kappa(\rpa,\rpe)k j_l(k\rpa)\exp(-i\bl \cdot \rpe)  .
\label{Flat_Fou}
\ee

\n
In this expansion $\bl$ depicts a 2D angular wavenumber and $k$ represents a wavenumber in the radial direction. It will also be
advantageous in certain situations when a harmonic expansion is only performed on the surface of
the sky, but the radial dependence is kept in configuration space. As before we have assumed in the above expansion that Universe is flat.
Alternatively the eigenfunctions
 for the expansion needs to be suitably modified. The real space
correlation functions and their Fourier counterparts are related by the following expression:

\begin{equation}
\langle \kappa(\rpa_1,\rpe_1)\dots\kappa(\rpa_n,\rpe_n)\rangle_c = \int {d^2\bl_1\over 2\pi}\dots{d^2\bl_n\over 2\pi} \langle\kappa(\bl_1,\rpa_1)\dots\kappa(\bl_n,\rpa_n)\rangle_c
\exp[i(\bl\cdot\rpe_1+\cdots+\bl_n\cdot\rpe_n)].
\end{equation}

\n
The flat sky correlation hierarchy, which ensures translational symmetry of the 2D patch sky is expressed by the following
equations. The $r$ label is retained as we have not the performed the Fourier expansion
in the radial direction.

\begin{eqnarray}
&& \langle \kappa(\bl_1,\rpa_1)\kappa(\bl_2,\rpa_2)\rangle_c = \2p\delta_{2D}(\bl_1+\bl_2){\cal P}(l_1,\rpa_i) \\
&& \langle \kappa(\bl_1,\rpa_1)\kappa(\bl_2,\rpa_2)\kappa(\bl_3,\rpa_3)\rangle_c = \2p\delta_{2D}(\bl_1+\bl_2+\bl_3){\cal B}_3(l_1,l_2,l_3;\rpa_i) \\
&& \langle \kappa(\bl_1,\rpa_1)\kappa(\bl_2,\rpa_2)\kappa(\bl_3,\rpa_3)\kappa(\bl_4,\rpa_4)\rangle_c = \2p\delta_{2D}(\bl_1+\bl_2+\bl_3+\bl_4){\cal T}_4(l_1,l_2,l_3,l_4;\rpa_i).
\end{eqnarray}

\n
The labels $r_i$ which appears as the arguments for multispectra denote all the radial coordinates which are involved
in their definition, e.g. $r_i =r_1,\dots,r_3$ for the bispectrum. The treatment for trispectra is more complicated
as it also gets disconnected Gaussian contributions (which are non-zero even in the absence of
any non-Gaussianity) and the contribution from the reduced segment, discussed above, which carries all the information about non-Gaussianity
at the level of fourpoint.

\ben
&& \langle \kappa(\bl_1,\rpa_1)\kappa(\bl_2,\rpa_2)\kappa(\bl_3,\rpa_3)\kappa(\bl_4,\rpa_4)\rangle_G =
\2p \delta_{2D}(\bl_1+\bl_2) \2p \delta_{2D}(\bl_3+\bl_4){\cal P}(l_1,\rpa_1){\cal P}(l_3,\rpa_3) \nn \\
&& \qquad \qquad + \2p \delta_{2D}(\bl_1+\bl_3) \2p \delta_{2D}(\bl_2+\bl_4){\cal P}(l_1,\rpa){\cal P}(l_2,\rpa)
+ \2p \delta_{2D}(\bl_1+\bl_4) \2p \delta_{2D}(\bl_2+\bl_3){\cal P}(l_1,\rpa_1){\cal P}(l_4,\rpa_4) \nn \\
&& \langle \kappa(\bl_1,r_1)\kappa(\bl_2,r_2)\kappa(\bl_3,\rpa_3)\kappa(\bl_4,\rpa_4)\rangle_c =
\2p\delta_{2D}(\bl_1+\bl_2+\bl_3+\bl_4){\cal T}_4(l_1,l_2,l_3,l_4,L;\rpa_i).
\een


\n
Following the discussion in \cite{Hu99} and \cite{OkaHu02} we write the trispectra $T^{l_1l_2}_{l_3l_4}$ in terms
of the reduced trispectra as follows:

\begin{equation}
{\cal T}^{l_1l_2}_{l_3l_4} = {\cal R}^{l_1l_2}_{l_3l_4}(l_{12}) + {\cal R}^{l_1l_3}_{l_2l_4}(l_{13}) + {\cal R}^{l_1l_4}_{l_3l_2}(l_{14}).
\end{equation}

\n
We have used the notation ${\bf l}_{12} = {\bf l}_1+{\bf l}_2$. We will quote the results obtained in \cite{Hu99} which relate
the multi-spectra defined in the full-sky analysis to their flat-sky counterpart.

\be
C_l(k;r) = {\cal P}(k;r); \qtwo  B_{l_1l_2l_3}(r_i) = I_{l_1l_2l_3}{\cal B}(l_1,l_2,l_3;r_i); \qtwo R^{l_1l_2}_{l_3l_4}(L,r_i) =
I_{l_1l_2L}I_{l_3l_4L}{\cal R}^{(l_1l_2)}_{(l_3l_4)}(L,r_i).
\ee

\n
While the results for the power spectrum and bispectrum are straightforward the reduced trispectrum is
more involved as it depends on the choice of the diagonal of the quadrilateral constructed out of the vectors $l_i$s.
The implemenation of the momentum conservation is imposed by the translational symmetry is built in the definition
is

\ben
\langle \kappa(\bl_1,\rpa_1)\kappa(\bl_2,\rpa_2)\kappa(\bl_3,\rpa_3)\kappa(\bl_4,\rpa_4)\rangle_c =
\2p\int \delta_{2D}(\bl_1+\bl_2+\bl) \delta_{2D}(\bl_3+\bl_4-\bl){\cal T}_4(l_1,l_2,l_3,l_4;l,\rpa_i) \qthree \nn \\
 =\int \left [ \delta_{2D}(\bl_1+\bl_2+\bl) \delta_{2D}(\bl_3+\bl_4-\bl){\cal R}^{l_1l_2}_{l_3l_4}(l) +
\delta_{2D}(\bl_1+\bl_3+\bl) \delta_{2D}(\bl_2+\bl_4-\bl){\cal R}^{l_1l_3}_{l_2l_4}(l) +
\delta_{2D}(\bl_1+\bl_4+\bl) \delta_{2D}(\bl_2+\bl_3-\bl){\cal R}^{l_1l_4}_{l_2l_3}(l) \right ]d^2 \bl.
\een

The flat patch wave numbers $(l_i)$ are used within the parentheses which appear on the r.h.s. of the
equations, whereas their all-sky versions appear in the l.h.s. without the parentheses.

The radial independence in our calculation can also be displayed by doing a Fourier transform in the radial direction.
The transformations from all-sky to the Fourier representation are given by the following expressions. We use
the same notations ${\cal B}_3$ or ${\cal T}_4$ in both representations.

\be
{\cal B}_3(l_1,l_2,l_3;r_i) = \left ( \sqrt{2 \over \pi} \right)^3 \int d\rpa_1 j_l(k_1\rpa_1) \dots \int d\rpa_3 j_l(k_3\rpa_3)
{\cal B}_3(l_1,l_2,l_3;k_i).
\label{bi_radial}
\ee

\n
A similar expression holds for the trispectrum:

\be
{\cal T}_4(l_1,\dots,l_4;r_i) = \left ( \sqrt{2 \over \pi} \right)^4 \int d\rpa_1 j_l(k_1\rpa_1) \dots \int d\rpa_3 j_l(k_3\rpa_3)
{\cal T}_4(l_1,\dots,l_4;k_i).
\label{tri_radial}
\ee

For a flat sky we can work with various representations for the multispectra as before. The
real space variables ${\bf r} = (r_{\parallel},{\bf r}_{\perp})$ and their Fourier
representations using variables ${\bf k} = (k, {\bf l})$ are both useful.

\ben
&& \langle \kappa(k_1,\bl_1)\kappa(k_2,\bl_2)\rangle = (2\pi)^3 \delta_{2D}(\bl_1+\bl_2)\delta_{1D}(k_1+k_2){\cal P}(k_1,\bl_1) \\
&& \langle \kappa(k_1,\bl_1) \cdots \kappa(k_3,\bl_3) \rangle = (2\pi)^3 \delta_{2D}(\bl_1+\dots+\bl_2)\delta_{1D}(k_1+\dots+k_2) {\cal B}_3(k_i,\bl_i) \\
&& \langle \kappa(k_1,\bl_1) \cdots \kappa(k_4,\bl_4)\rangle = (2\pi)^3 \delta_{2D}(\bl_1+\dots+\bl_4)\delta_{1D}(k_1+\dots+k_2) {\cal T}_4(k_i,\bl_i).
\een

\n
The relations that we will be using most in our derivations are the orthogonality relationship of the Bessel functions
Eq.(\ref{eq:limber_approx1}) and the representation of the 2D Dirac delta function

\be
\int j_l(l_1 \rpa) j_l(l_2 \rpa) d\rpa = \left ( {\pi \over 2\rpa^2} \right ) \delta_{1D}(l_1 - l_2); \qquad
\int \exp\{i{\rpe} \cdot ({\bf l}_1 - {\bf l}_2)\} ~d^2 \rpe = \2p~\delta_{2D}({\bf l}_1 - {\bf l}_2).
\ee

\n
Our convention for the Fourier transform of an arbitrary  real-space  function $\kappa(\rpa,\rpe)$ to Fourier space $\kappa(k,\bl)$
for small angular scale approximation is given by:

\be
\kappa(k,\bl) =   \sqrt{2\over \pi}\int{d\rpa} \int {d^2\rpe \over {2\pi}} k j_l(k\rpa)\exp(-\bl\cdot\rpe) \kappa (\rpa,\rpe); \qtwo
\kappa(\rpa,\rpe) =  \sqrt{2\over \pi}\int {d k} \int {d^2 {\bf l} \over {2\pi}}  k j_l(k\rpa)\exp(-\bl\cdot\rpe) \kappa (k,\bl).
\ee

\n
We will also be working with partial transforms such as $\kappa(k,\rpe)$ and $\kappa(\rpa,\bl)$  which are defined in an obvious way.

\subsection{Flat Sky Without Mask}

\n
In this section we will consider the power spectra associated with multi-spectra in a flat patch of
sky suitable for smaller surveys.
 These are cross power spectra $P_{pq}(l)$, that are the Fourier transforms of cross correlation function
of fields constructed from the moments of the original field $\kappa(\br)$, e.g. $\kappa^p(\br)$ and $\kappa^q(\br)$.
Being collapsed multipoint statistics, they carry information about the associated multispectra of order $p+q$, although they themselves are just two-point
objects in real space. We will be using the same convention for the Fourier transform as the previous section:

\be
\kappa(\rpe) = {1 \over 2\pi} \int \kappa({\bl}) \exp(i\bl\cdot\rpe)d^2\bl; \qquad
 \kappa(\bl) = {1 \over 2\pi} \int \kappa({\bl}) \exp(-i\bl\cdot\rpe)d^2\rpe; \qquad
 \langle \kappa(\bl_1) \kappa(\bl_2)\rangle_c = \2p {\cal P}(l)\delta_{2D}(\bl_1-\bl_2)
\ee

\n
In these expressions we have suppressed the explicit radial dependence - which will be introduced at the end of sections.
We will first consider the skew spectrum. The cumulant correlator of interest in this case is $\langle \kappa^2(\br_1)\kappa(\br_2)\rangle_c$.
This is related to the underlying bispectrum $B(l_1,l_2,l_3)$ and we denote the associated power spectrum
by $P_{21}(l)$.

\ben
&& \kappa^{(2)}(\bl) = {1 \over \2p} \int [\kappa^2({\rpe})] \exp(-i\bl\cdot\rpe)d^2\rpe; \qquad \kappa^{2}(\rpe) =  {1 \over \2p}\int \kappa(\bl_1)\kappa(\bl_2) \exp\{ (\bl_1 +\bl_2 - \bl) \cdot \rpe \}d^2\bl_1d^2\bl_2;  \\
&& \kappa^{(2)}(\bl) =  {1 \over \2p}\int \kappa(\bl_1) \kappa(\bl_2) \delta_{2D}(\bl_1+\bl_2-\bl) d^2\bl_1 d^2\bl_2.
\een

\n
Using the above expressions we can write down the flat-sky version of the skew spectrum:

\be
 {\cal P}_{21}(l)\equiv \langle \kappa^{(2)}(\bl)\kappa^*(\bl)\rangle_c = \int {\cal B}_3(l_1,l_2,l) {\cal S}(l_1,l_2,l) l_1l_2\,dl_1 dl_2.
\ee

\n
In deriving this we have carried out the angular integrals $\phi_{\bl_1}$ and $\phi_{\bl_2}$ by using the following
equation \citep{Hiv02}:

\be
\int d\phi_{\bl_1} \int d\phi_{\bl_2} \delta_{2D}(\bl_1+\bl_2+\bl) = 2\pi {\cal S}(l_1,l_2,l),
\label{hivon_eq}
\ee
where
\be
{\cal S}(l_1,l_2,l_3) \equiv {2 \over \pi} (l_1^2 + l_2^2 + l_3^2 - 2l_1l_2 - 2l_1l_3 - 2l_2l_3)^{-1/2}.
\label{eq.S}
\ee

\n
The derivation outlined above implicitly assumes that the bispectrum has no angular dependence in Fourier space. This is valid for the ``stellar'' model
we will be considering. Carrying through the analysis in a very similar way we can write down the first of a set of two degenerate power spectra
associated with the trispectrum $P_{22}(l)$ :

\be
{\cal P}_{22}(l) = \langle\kappa^{(2)}(\bl) \kappa^{(2)}(\bl)^*\rangle_c = \int \langle \kappa(\bl_1)\kappa(\bl_2)\kappa(\bl)\kappa(\bl_4)\rangle_c \delta_{2D}(\bl_1 + \bl_2 - \bl) \delta_{2D}(\bl_3+\bl_4+\bl)
d^2\bl_1 d^2\bl_2 d^2\bl_3 d^2\bl_4 .
\ee

\n
Using Eq.(\ref{hivon_eq}) again to simplify the angular integrals we find

\begin{equation}
{\cal P}_{22}(l) = \int l_1 dl_1 \int l_2 dl_2 \int l_3 dl_3 \int l_4 dl_4 {\cal T}_{4}(l_1,l_2,l_3,l_4){\cal S}(l_1,l_2,l){\cal S}(l_3,l_4,l).
\end{equation}

\n
In an analogous way, going through the same algebra for the other degenerate power spectra associated with trispectrum we find

\ben
&& \kappa^{(3)}(\bl)= \int \kappa^3(rpe)\exp(i\bl \cdot\rpe)d^2\bk; \qquad \kappa^{(3)}(\bl)= \int \kappa(\bl_1)\kappa(\bl_2)
\kappa(\bl_3)\delta_{2D}(\bl_1+\bl_2+\bl_3-\bl) d^2\bl_1 d^2\bl_2 d^2\bl_3  \\
&& P_{31}(l) = \langle \kappa^{(3)}(\bl)^* \kappa(\bl) \rangle = \int {\cal T}_4(l_1,l_2,l_3,l_4){\cal S}(l_1,l_2,l') {\cal S}(l',l_3,l) l_1 l_2 l_3 l' dl_1dl_2 dl_3 dl'.
\een

\n

In our derivation we have decomposed the $\delta_{2D}$ function $\delta_{2D}(\bl_1+\bl_2+\bl_3-\bl)$ in terms of two $\delta_{2D}$ function
$\delta_{2D}(\bl_1+\bl_2+\bl_3-\bl) = \int \delta_{2D}(\bl_1+\bl_2+\bl')\delta_{2D}(\bl_3-\bl-\bl')d^2\bl'$ and
used Eq.(\ref{hivon_eq}) individually on each of them. We assume no angular dependence for the trispectrum, valid for the stellar model that
we consider. For the radial dependence we need to replace the ${\cal T}_4(l_1,l_2,l_3,l_4)$ with  ${\cal T}_4(l_1,l_2,l_3,l_4;r_i)$
which will not affect the rest of the analysis.

The power spectra ${\cal P}_{31}(l)$ and ${\cal P}_{22}(l)$ both probe the
configuration dependence of the trispectrum ${\cal T}_4(l_1,l_2,l_3,l_4)$
in a restricted sense. While the power spectrum ${\cal P}_{22}(l)$ considers all possible configurations with the diagonal
of the quadrangle (formed by momentum vectors $\bl_i$) constant, ${\cal P}_{31}(l)$ gives an estimate where one side of the
quadrangle is kept fixed while all other sides as well as both diagonals vary.

Given a specific model for the multispectra now we can compute the associated power spectra and compare them with
simulated and observed data. To consider radial dependence we need to start our analysis from Eq.(\ref{bi_radial})
for the bispectrum and Eq.(\ref{tri_radial}) for the trispectrum. Following exactly similar analysis we will
recover the flat sky versions Eq.(\ref{sphere22}) and Eq.(\ref{sphere31}).

\subsection{Flat Sky With a Mask}

\n
We will consider a very general mask $w(\br)$ without enforcing any symmetry. We will show that the
estimated power spectra from a masked data is a convolved estimate of the underlying true
estimates. We will derive the convolution function and how it is determined by the properties
of the mask and develop procedures to deconvolve the effect of the mask to have an
unbiased estimator. The analysis will parallel our discussion for the spherical sky.
Let us start by defining a convolved scalar field $\tilde \kappa(\br)$ which depends both on the
original field $\kappa(\br)$ as well as the mask $w(\br)$, $\tilde \kappa(\br) = \kappa(\br)w(\br)$
In the Fourier domain this takes the form of a convolution. If we represent the Fourier transform
of $\tilde \kappa(\br)$ by $\tilde \kappa(\bl)$ we can express it in terms of the following expression
using the Fourier transform of the mask $w(\bl)$ and the unsmoothed  convergence field $\kappa(\bl)$:

\be
\tilde \kappa(\bl) = \int \kappa(\bl_1) w(\bl_2) \delta_{2D}(\bl_1+\bl_2-\bl) d^2\bl_1 d^2\bl_2,
\ee

 \n
which with the help of a kernel $K_{\bl\bl'}[w]$, which encodes the effect of the mask,  takes a more compact form:

\be
\tilde \kappa(\bl_1) = \int d^2 \bl_2 K_{\bl_1\bl_2} [w]; \qtwo K_{\bl_1\bl_2}[w] \equiv \int d^2\bl_3 w(\bl_3) \delta_{2D}(\bl_1-\bl_2+\bl_3).
\ee

\n
The power spectrum associated with the masked fields is given in terms of the
kernel $S(l,l_1,l_2)$ and the power spectrum of the unmasked field  $P(l_1)$
and the power spectrum of the mask $P_w(l) = {1 \over 2\pi} \int d\phi_{\bl} w(\bl) w(\bl)^*$ \citep{Hiv02}:

\be
\tilde {\cal P}_{11}(l) = {1 \over 2\pi} \int d\phi_{\bl} \langle \tilde\kappa(\bl)\tilde\kappa^*(\bl)\rangle =  \int {\cal P}(l_1) P_w(l_2)
{\cal S}(l,l_1,l_2)l_2 dl_2 l_1dl_1.
\ee

\n
We can rewrite the same equation in a compact form \citep{Hiv02}:

\be
\tilde {\cal P}_{11}(l_2) = \int {\cal P}(l_1) M_{l_1l_2}  l_1 dl_1; \qtwo M_{l_1l_2} = 2\pi \int P_w(l) {\cal S}(l,l_1,l_2) ldl.
\ee

\n
Here we have introduced the kernel $M_{\bl_1\bl_2}$ and used the notation $d^2\bl = ldld\phi$ for the integration variables
on the surface of the sky (2D flat patch). $S(l_1,l_2,l_3)$ is defined in Eq.(\ref{eq.S}). This result has general applicability
and does not depend on any specific mask properties. Hence $\tilde P_{11}(l)$ can be used as an estimator for the deconvolved power spectrum  $P(l_1)$.
Carrying out a exactly similar procedure we can obtain the results for the
skew spectrum or the kurt-spectrum. If we denote the power
spectrum associated with the correlation function $\langle \tilde \kappa^p(\br_1)\tilde \kappa^q(\br_2)\rangle_c$
by $\tilde P_{pq}(l)$ and the deconvolved power spectrum $P_{pq}(l)$ which is the Fourier
representation of the correlation function $\langle \kappa^p(\br_1) \kappa^q(\br_2)\rangle_c$
then they are related by the same expression:

\be
\tilde {\cal P}_{pq}(l_1) = {1 \over 2\pi} \int d\phi_{\bl_1} \langle\tilde\kappa(\bl_1)^p\tilde\kappa^*(\bl_1)^q\rangle  =
\int M_{l_1l_2} {\cal P}_{pq}(l_2) l_2 dl_2.
\ee

\n
The Fourier transforms of the squared field $\kappa^2$ with
a mask are related by the usual coupling matrix,
\ben
\tilde \kappa^{(2)}(l_2) = \int K_{\bl_1\bl_2}[w] \kappa^{(2)}(l_1) d^2\bl_1; \qquad
\tilde \kappa^{(3)}(l_2) = \int K_{\bl_1\bl_2}[w] \kappa^{(3)}(l_1) d^2\bl_1.
\een

\n
and masked power spectra $\tilde {\cal P}_{21}(l_1) = \langle \tilde \kappa^{(2)}(\bl_1)\tilde\kappa(\bl_2)\rangle = \2p\delta_{2D}(\bl_1-\bl_2)$  are given by
\be
\tilde {\cal P}_{21}(l_1) = \int M_{l_1l_2} {\cal P}_{21}(l_2) l_2 dl_2; \qtwo
\tilde {\cal P}_{22}(l_1) = \int  M_{l_1l_2} {\cal P}_{22}(l_2) l_2 dl_2;  \qtwo
\tilde {\cal P}_{31}(l_1) = \int M_{l_1l_2} {\cal P}_{31}(l_2) l_2 dl_2.
\ee

\n
These equations represent the flat-sky version of the all-sky expressions Eq.(\ref{partsphere22}) and Eq.(\ref{partsphere31}). They generalise the results
obtained for the flat-sky power spectrum by \cite{Hiv02}. The coupling matrix
$M_{l_1l_2}$ introduced in this section is the flat-sky analogue of its all-sky counterpart. The power spectra described here, i.e.  ${\cal P}_{11}(l),{\cal P}_{21}(l),{\cal P}_{22}(l), {\cal P}_{31}(l)$
are associated with relevant multispectra of the same order. These are useful probes of associated
mutispectra as they do not compress the available information to a single number and retain some
of the relevant shape dependence. The fact that unbiased estimators can be constructed by simple inversion
means estimation of such multispectra from simulations and observational data may be realistically possible
even in the presence of complicated masks with non-trivial topology. The issues of analysis of noise
subtraction can be dealt with in a similar manner.

\section{Conclusions}

Future weak lensing surveys will play a big part in further
reducing the uncertainty in fundamental parameters, including those
that describe the evolution of equation of state of dark energy \cite{Euclid}.
Weak lensing surveys can exploit both the angular diameter distance
and the growth of structure to constrain cosmological parameters,
and can test the gravity model \cite{HKV07,Amendola08,Benyon09}.   For recent results, see \cite{Schrabback09, Kilbinger09}.  
Such constraints from weak lensing
are complementary to those obtained from cosmic microwave background
studies and from galaxy surveys as they probe structure formation in
the dark sector at a relatively low redshift range. Initial studies
in weak lensing were restricted to studying two-point functions in
projection for the entire source distribution. It was, however,
found that binning sources in a few photometric redshift bins
can improve the constraints\cite{Hu99}. More recently a full 3D formalism has been
developed which uses photometric redshift of all sources without any
binning\cite{Heav03,Castro05,HKT06}. These studies have demonstrated that 3D lensing can provide
more powerful and tighter constraints on the dark energy equation of
state parameter, on neutrino masses \cite{deBernardis09}, as well as testing braneworld
and other alternative gravity models. Most of these 3D works have
primarily focussed on power spectrum analysis, but in future
accurate higher-order statistic measurement should be possible (e.g. \cite{TakadaJain04, Semboloni09}.

In this paper we have generalized such studies analytically to
multi-spectra which takes us beyond conventional power spectrum
analysis. The previously obtained analytical results were developed
for the statistical study of weak lensing observable using generic
models for the multispectra of the underlying mass distribution.
Later on we specialize the results for the case of specific examples
using the hierarchical ansatz, where higher-order multispectra are
constructed from various products of power spectra organized in all
possible topological diagrams with different amplitudes. The
analytical results are developed both for near all-sky surveys as
well as for flat patches of the sky. The formalism developed does
not depend on the background cosmology and can be used to predict
level of non-Gaussianity for both primary as well as secondary
non-Gaussianity.

The higher-order multispectra contain a wealth of information in
through their shape dependence. Though partly degenerate, this
information can be invaluable for constraining structure formation
scenarios. However determination of the multispectra and their
complete shape dependence is not an easy task from noisy data. In
this paper we advocate a set of statistics called ``cumulant correlators''
which were first used in real space in the context of galaxy surveys
and later extended to CMB studies. Here we have presented a general
formalism for the study of the power spectra or the Fourier
transforms of these correlators. We present a 3D analysis which
takes into account the radial as well as on the surface of the sky
decomposition. We start by relating various representation of
multispectra in three dimensions. We relate the spherical representation and
the Fourier representations with other possibilities: mixed modes of
representations. These allow us to relate the harmonic decomposition
of convergence directly with that of underlying mass distribution.

We have restricted this study to the third and fourth order, though
it can be generalised to higher order and some of our results are
valid at arbitrary order. At third order, we define a power spectrum
which compresses information associated with a
bispectrum to a power spectrum. This power spectrum
$C_l^{2,1}(r_2,r_1)$ is the cross-power spectrum associated with
squared convergence maps $\kappa^2(r_1,\oh)$ constructed at a
specific radial distance $r_1$ against  $\kappa^2(r_2,\oh)$ at $r_2$
In a similar manner we also associate power spectra $C_l^{2,2}$ and
$C_l^{3,1}$ with associated trispectra $T^{l_1l_2}_{l_3l_4}(L;r_i)$.
There are two different power spectra at the level of trispectra
which are related to the respective real-space correlation functions
$\langle \kappa^2(r_1,\oh)\kappa^2(r_2,\oh')\rangle$ and $\langle
\kappa^3(r_1,\oh)\kappa(r_2,\oh')\rangle$. We expressed these
real-space correlators in terms of their Fourier space analogue
which take the form of $C_l^{3,1}(r_2,r_1)$ and
$C_l^{2,2}(r_2,r_1)$. We develop analytical expressions to take into
account the photometric redshift errors in these power spectra.
While we present formalisms which are completely general, we also
use the Limber approximation to reduce the dimensionality of the
relevant integrations. These when combined with specific
hierarchical models of gravitational clustering can make analytical
results remarkably simpler.

The statistics presented here will be a useful tool in studying non-Gaussianity in
alternative theories of gravity; which are one of the important
science drivers for the future generations of weak lensing surveys. We plan to
present detailed results elsewhere in future.

\section{Acknowledgements}
\label{acknow} Initial phase of this work was completed when DM was
supported by a STFC rolling grant at the Royal Observatory,
Institute for Astronomy, Edinburgh. DM also acknowledges support
from STFC standard grant ST/G002231/1 at the School of Physics and
Astronomy at Cardiff University where this work was completed. It is
a pleasure to thank Asantha Cooray and Patrick Valageas for many
useful discussions.

\bibliography{paper.bbl}

\begin{thebibliography}{}

\bibitem[\protect\citeauthoryear{Amendola, Kunz \& Sapone}{2008}]{Amendola08}
Amendola, L., Kunz M., Sapone D., 2008, JCAP, 04, 13

\bibitem[\protect\citeauthoryear{Barber, Munshi \& Valageas}{2004}]{BaMuVa04}
Barber A.J., Munshi D., Valageas P., 2004, MNRAS, 347, 667

\bibitem[\protect\citeauthoryear{Bartolo et al}{2004}]{BART04}
Bartolo N., Komatsu E., Matarrese S., Riotto A., 2004, Phys.Rept., 402, 103

\bibitem[\protect\citeauthoryear{Beacon, Refregier \& Ellis }{2000}]{BRE00}
Beacon D.J., Refregier A., Ellis R.S., 2000, MNRAS, 318,625

\bibitem[\protect\citeauthoryear{Benyon, Bacn \& Koyama}{2009}]{Benyon09}
Benyon E., Bacon D.J., Koyama K., 2009, astroph/0910.1480

\bibitem[\protect\citeauthoryear{Bernardeau \& Schaeffer}{1992}]{BerSch92}
Bernardeau F., Schaeffer R., 1992, A\&A, 255, 1

\bibitem[\protect\citeauthoryear{Bernardeau \& Valageas}{2000}]{BerVal00}
Bernardeau F., Valageas P., 2000, A\&A, 364, 1

\bibitem[\protect\citeauthoryear{Bernardeau, Van Waerbeke \& Mellier}{1997}]{BerVanMell97}
Bernardeau F., Van Waerbeke L., Mellier Y., 1997, A\&A, 322, 1

\bibitem[\protect\citeauthoryear{Bernardeau, Mellier \& Van Waerbeke}{2002}]{BerVanMell02}
Bernardeau F., Mellier Y., Van Waerbeke L., 2002, A\&A, 389, L28

\bibitem[\protect\citeauthoryear{Bernardeau, Mellier \& va Waerbeke}{2003}]{berMellwaer02}
Bernardeau F., Mellier Y. van Waerbeke L., 2003, A\&A, 389, L28

\bibitem[\protect\citeauthoryear{Bernardeau, van Waerbeke \&  Mellier}{2003}]{berludoMell03}
Bernardeau F., van Waerbeke L., Mellier Y., 2003, A\&A, 397, 405

\bibitem[\protect\citeauthoryear{Bernardeau et al}{2002}]{Bernardreview02}
Bernardeau F., Colombi S., Gaztanaga E., Scoccimarro R., 2002, Phys.Rept.,367, 1

\bibitem[\protect\citeauthoryear{Castro et al}{2005}]{Castro05}
Castro P.G., Heavens A.F., Kitching T.D., 2005, Phys.Rev. D72, 023516

\bibitem[\protect\citeauthoryear{Coles, Melott \& Munshi}{1999}]{CMM99}
Coles P., Melott A.L., Munshi D., 1999, ApJ, 521, L5

\bibitem[\protect\citeauthoryear{Cooray}{2005}]{Cooray01}
Cooray A, 2001, Phys.Rev. D, 64, 043516

\bibitem[\protect\citeauthoryear{Cooray \& Seth}{2002}]{CooSeth02}
Cooray A., Seth R., 2002, Phys. Rep. 372, 1

\bibitem[\protect\citeauthoryear{Cooray}{2006}]{Cooray06}
Cooray A, 2006, PRL, 97, 261301

\bibitem[\protect\citeauthoryear{Cooray, Li \& Melchiorri}{2003}]{CooLiMel08}
Cooray A., Li C., Melchiorri A., 2008, Phys.Rev.D, 77,103506


\bibitem[\protect\citeauthoryear{Creminelli et al.}{2006}]{Crem06}
Creminelli P., Nicolis A., Senatore L., Tegmark M., Zaldarriaga M., 2006, JCAP, 5, 4

\bibitem[\protect\citeauthoryear{de Bernardis et al.}{2009}]{deBernardis09}
de Bernardis F., Kitching T.~D., Heavens, A., Melchiorri, A., 2009, Phys. Rev. D80, 123509

\bibitem[\protect\citeauthoryear{Fry}{1984}]{Fry84}
Fry J.N., 1984, ApJ, 279, 499

\bibitem[\protect\citeauthoryear{Heavens}{2003}]{Heav03}
Heavens A.F., 2003, MNRAS, 343, 1327

\bibitem[\protect\citeauthoryear{Heavens et al.}{2000}]{HRH00}
Heavens A.~F., Refregier A., Heymans C.E., 2000, MNRAS, 319, 649

\bibitem[\protect\citeauthoryear{Heavens et al}{2006}]{HKT06}
Heavens A.~F., Kitching T.~D., Taylor A.N., 2006, MNRAS, 373, 105

\bibitem[\protect\citeauthoryear{Heavens, Kitching \& Verde}{2007}]{HKV07}
Heavens A.~F., Kitching T.~D., Verde L., 2007, MNRAS, 380, 1029

\bibitem[\protect\citeauthoryear{Hivon et al.}{2002}]{Hiv02} 
Hivon E., G{\'o}rski K.~M., Netterfield C.~B., Crill B.~P., Prunet S., 
Hansen F., 2002, ApJ, 567, 2 

\bibitem[\protect\citeauthoryear{Hoek, Yee \& Gladders}{2002}]{HoekYeeGlad02}
Hoekstra H., Yee H.~K.~C., Gladders M. D., 2002, ApJ, 577, 595

\bibitem[\protect\citeauthoryear{Hu}{1999}]{Hu99}
Hu W., ApJ., 1999, 522, L21

\bibitem[\protect\citeauthoryear{Hui}{1999}]{Hui99}
Hui L., ApJ.,1999, 519, L9

\bibitem[\protect\citeauthoryear{Jain, Seljak \& White}{2000}]{JSW00}
Jain B, Seljak U., White S. Astrophys.J., 2000, 530, 547

\bibitem[\protect\citeauthoryear{Jain \& Seljak}{1997}]{JainSeljak97}
Jain B., Seljak U., 1997, ApJ, 484, 560

\bibitem[\protect\citeauthoryear{Kaiser}{1992}]{Kaiser92}
Kaiser N. 1992. ApJ, 388, 272

\bibitem[\protect\citeauthoryear{Kaiser, Wilson \& Luppino}{2000}]{KWL00}
Kaiser N., Wilson G., Luppino G.A., astro-ph/0003338 


\bibitem[\protect\citeauthoryear{Kilbinger et al.}{2009}]{Kilbinger09}
Kilbinger M., et al., 2009, A\& A, 497, 677

\bibitem[\protect\citeauthoryear{Kitching et al.}{2008}]{Kit08}
Kitching T.D., Heavens A. F., Verde L., Serra P.,  Melchiorri A., 
Phys.Rev. 2008, D77, 103008

\bibitem[\protect\citeauthoryear{Limber}{1954}]{Limb54}
Limber D.N., 1954, ApJ, 119, 665

\bibitem[\protect\citeauthoryear{LoVerde \& Afshordi}{2008}]{LoAf08}
LoVerde M., Afshordi N. 2008, Phys.Rev.D78, 123506

\bibitem[\protect\citeauthoryear{Massey et al}{2007}]{Massey07}
Massey R. et al, 2007, Nature,  445, 286

\bibitem[\protect\citeauthoryear{Massey et al}{2007a}]{Massey07a}
Massey et al, 2007, ApJS, 172, 239

\bibitem[\protect\citeauthoryear{Munshi}{2000}]{Mu00} 
Munshi D., 2000, MNRAS, 318, 145 

\bibitem[\protect\citeauthoryear{Munshi et al}{1999}]{MuBaMeSch99}
Munshi D., Bernardeau F., Melott A.L., Schaeffer R.,1999, MNRAS, 303, 433

\bibitem[\protect\citeauthoryear{Munshi \& Coles}{2000}]{MuCo00} 
Munshi D., Coles P., 2000, MNRAS, 313, 148 

\bibitem[\protect\citeauthoryear{Munshi \& Coles}{2002}]{MuCo02} 
Munshi D., Coles P., 2002, MNRAS.329, 797 

\bibitem[\protect\citeauthoryear{Munshi \& Coles}{2003}]{MuCo03} 
Munshi D., Coles P., 2003, MNRAS, 338, 846

\bibitem[\protect\citeauthoryear{Munshi, Coles \& Melott}{1999a}]{MuCoMe99a} 
Munshi D., Coles P., Melott A.L., 1999a, MNRAS, 307, 387

\bibitem[\protect\citeauthoryear{Munshi, Coles \& Melott}{1999b}]{MuCoMe99b} 
Munshi D., Coles P., Melott A.L., 1999b, MNRAS, 310, 892

\bibitem[\protect\citeauthoryear{Munshi \& Heavens}{2009}]{MuHe09} 
Munshi D., Heavens A., MNRAS (in press)

\bibitem[\protect\citeauthoryear{Munshi et al.}{2009}]{Munshi_kurt}
Munshi D. et al. 2009 arXiv:0910.3693

\bibitem[\protect\citeauthoryear{Munshi \& Jain}{2000}]{MuJa00} 
Munshi D., Jain B., 2000, MNRAS, 318, 109 

\bibitem[\protect\citeauthoryear{Munshi \& Jain}{2001}]{MuJai01} 
Munshi D., Jain B., 2001, MNRAS, 322, 107 

\bibitem[\protect\citeauthoryear{Munshi, Melott \& Coles}{1999}]{MuMeCo99} 
Munshi D., Melott A.L., Coles P., 1999, MNRAS, 311, 149

\bibitem[\protect\citeauthoryear{Munshi \& Valageas}{2005}]{MuVa05} 
Munshi D., Valageas P., 2005, RSPTA, 363, 2675 

\bibitem[\protect\citeauthoryear{Munshi, Valageas \& Barber}{2004}]{MuVaBa04} 
Munshi D., Valageas P., Barber A.~J., 2004, MNRAS, 350, 77 

\bibitem[\protect\citeauthoryear{Munshi et al.}{2008}]{MuPhysRep08} 
Munshi D., Valageas P., van Waerbeke L., Heavens A., 2008, PhR, 462, 67 

\bibitem[\protect\citeauthoryear{Okamoto \& Hu}{2002}]{OkaHu02}
Okamoto T, Hu W., Phys.Rev. 2002, D66, 063008

\bibitem[\protect\citeauthoryear{Refregier et al.}{2010}]{Euclid}
Refregier et al., 2010, astroph/1001.0061

\bibitem[\protect\citeauthoryear{Schaeffer}{1984}]{Schaeffer84} 
Schaeffer R., 1984, A\&A, 134, L15

\bibitem[\protect\citeauthoryear{Schrabback et al.}{2009}]{Schrabback09}
Schrabback et al, 2009, astro.co 0911.0053

\bibitem[\protect\citeauthoryear{Schneider et al}{2002}]{Schneider98}
Schneider P., Van Waerbeke L., Jain B., Kruse G., 1998, MNRAS, 296, 873, 873

\bibitem[\protect\citeauthoryear{Schrabback et al.}{2009}]{Schrabback09}
Schrabback et al., 2009, astroph/0911.0053

\bibitem[\protect\citeauthoryear{Scoccimarro et al}{1998}]{Scocci98} 
Scoccimarro R. et al,Astrophys.J. 1998, 496 586

\bibitem[\protect\citeauthoryear{Semboloni et al}{2008}]{Semboloni08}
Semboloni et al, 2008, MNRAS, 388, 991

\bibitem[\protect\citeauthoryear{Semboloni et al}{2009}]{Semboloni09}
Semboloni E., Tereno I., van Waerbeke L, Heymans C.,  2009, MNRAS, 397, 608

\bibitem[\protect\citeauthoryear{Smith \& Zaldarriaga}{2006}]{SmZa06} 
Smith K.~M., Zaldarriaga M., 2006, arXiv:astro-ph/0612571

\bibitem[\protect\citeauthoryear{Stebbins}{1996}]{Stebbins96}
Stebbins A., arXiv:astro-ph/9609149

\bibitem[\protect\citeauthoryear{Szapudi \& Szalay}{1993}]{SzaSza93}
Szapudi I., Szalay A.S., 1993, ApJ, 408, 43

\bibitem[\protect\citeauthoryear{Szapudi \& Szalay}{1997}]{SzaSza97}
Szapudi I., Szalay A.S., 1997, ApJ, 481, L1

\bibitem[\protect\citeauthoryear{Takada \& Jain}{2004}]{TakadaJain04} 
Takada M., Jain B.,  MNRAS, 348 (2004) 897

\bibitem[\protect\citeauthoryear{Takada \& Jain}{2003}]{TakadaJain03} 
Takada M., Jain B.,  2003, MNRAS, 344, 857

\bibitem[\protect\citeauthoryear{Takada \& White}{2003}]{TakadaWhite03}
Takada M., White M.,2001, ApJ. 601, L1

\bibitem[\protect\citeauthoryear{Takada \& Jain}{2009}]{TakadaJain09}
Takada M. Jain B., 2009, MNRAS, 395, 2065 

\bibitem[\protect\citeauthoryear{Valageas}{2000}]{Valageas00}
Valageas P., 2009, A\&A, 356, 771

\bibitem[\protect\citeauthoryear{Valageas, Munshi \& Barber}{2005}]{VaMuBa05} 
Valageas P., Munshi D., Barber A.~J., 2005, MNRAS, 356, 386 

\bibitem[\protect\citeauthoryear{Valageas \& Munshi}{2004}]{vaMu04} 
Valageas P., Munshi D., 2004, MNRAS, 354, 1146 

\bibitem[\protect\citeauthoryear{Valageas, Barber, \& Munshi}{2004}]{VaMuBa04} 
Valageas P., Barber A.~J., Munshi D., 2004, MNRAS, 347, 654 

\bibitem[\protect\citeauthoryear{Waerbeke et al}{2000}]{Waerbeke00}
van Waerbeke L. et al. 2000, A\&A, 358, 30

\bibitem[\protect\citeauthoryear{Waerbeke et al}{2002}]{Waerbeke02}
van Waerbeke L. et al. 2002, A\&A, 393, 369

\bibitem[\protect\citeauthoryear{Wittman et al}{2000}]{Wittman00}
Wittman D. et al. 2000, Nature, 405, 143

\end{thebibliography}

\appendix

\section{Realistic selection function and Photometric Redshift Errors}
\label{photo_z}

The results obtained in the main text was simplified for clarity, ignoring the fact that in a realistic survey, the average number density of sources will decline with distance, and the distances estimated from photometry will include errors.   We consider these here.

\subsection{All Sky results}

The lensing potential can only be sampled at the position of galaxies. Hence it can be written as
sum over galaxy positions. This discrete sum can be expressed as

\begin{equation}
\kappa_{lm}^O(k;r) = \sqrt{2 \over \pi } \sum_g \kappa(\br)k j_l(kr_g^0) Y_{lm}(\ho_g)W(r_g^0).
\end{equation}

\n
Here $W$ is an arbitrary weight function, and $r_g^0$ is the distance to galaxy $g$ assuming a fiducial cosmology.  The convergence depends of course on teh correct distance in the true cosmology, $r$.  Replacing the discrete sum over the galaxy positions with an integral we can write

\begin{equation}
 \kappa_{lm}^O(k;r) = \sqrt{2 \over \pi } \int d^3\br^0 ~n(r)~  \kappa(\br)k j_l(kr_g^0) Y_{lm}(\ho_g)w(r_g).
\end{equation}

\n
The quantity $n(\br)$ comprised of sum of delta functions which peaks at observed positions of the
galaxies.  Ensemble averaging of these quantities will reduce the equation to

\begin{equation}
 \kappa_{lm}^O(k;r) = \sqrt{2 \over \pi } \int d^3~\br^0 {\bar n(r)}  \kappa(\br)k j_l(kr^0) Y_{lm}(\ho_g)W(r^0)
\end{equation}

\n
Because of the discrete nature of source galaxies the estimator will have a scatter due
to the shot noise. It will also have contribution from source clustering.
While we will investigate the effects of photometric redshift errors, we will ignore the uncertainties in the
photometric redshift distribution of sources which means we can write  $n(r^0)d^3r^0 = \bar n_z(z_p)dz_p d\ho /4\pi$.
We will ignore the effect of source clustering which does not play a dominant role in the error budget for deep surveys.

\begin{equation}
\kappa_{lm}^O(k;r) = \sqrt{1 \over 8\pi^3 } \int dz ~d\ho~  {\bar n_z(z_p)}  \kappa(\br)k j_l(kr^0) Y_{lm}(\ho)W(z_p)
\end{equation}
where $r^0$ is the fiducial distance at redshift $z_p$.

The primary effect of photometric redshifts is to smooth the source distributions along the line of sight
distribution. If $p(z|z_p)$ denotes the probability of the true redshift being $z$ given the photometric redshift  $z_p$, the above equation, when modified to take into account the effect of photometric redshift error can be written as:

\begin{equation}
\kappa_{lm}^O(k;r) = \sqrt{1 \over 8\pi^3 } \int dz_p \int dz \int d\ho  {\bar n_z(z_p)} p(z|z_p)  \kappa(\br)k j_l(kr^0) Y_{lm}(\ho)W(z_p)
\end{equation}

\n
After expanding the $\kappa(\br)$ and carrying out the angular integrations we can eventually arrive at the following expression:

\be
\kappa_{lm}^O(k;r) = \sqrt{1 \over 8\pi^3 } \int dz_p \int dz {\bar n_z(z_p)} p(z|z_p) k j_l(kr^0) \int dk' k' j_l(k'r^0)W(z_p)
\kappa_{lm}(k';r)
\ee

\n
Typically $p(z|z_p)$ is modelled as a Gaussian for simplicity, though it may have catastrophic failures.  These can be included by modification of $p(z|z_p)$.

\begin{equation}
p(z|z_p) = { 1 \over \sqrt {2 \pi} \sigma_z(z) } \exp \left [ { -(z_p- z + z_{bias})^2 \over 2 \sigma_z^2(z)}\right ].
\end{equation}

\n
In this expression $z_{bias}$ is the possible bias in the photometric redshift calibration. The dispersion in error $\sigma_z(z)$ depends on
the redshift. The photometric redshift errors evidently introduces error in the radial direction.
Having expressed the $\kappa^O_{lm}(k;r)$ by taking into account the photometric redshift errors in terms of $\kappa_{lm}(k,r)$, we can now
construct the multispectra for the observed harmonics by relating $\kappa_{lm}(k;r)$ to $\delta_{lm}(k;r)$ as outlined in the main text .

\subsection{Flat Sky Expressions}

\n
Here we give flat-sky results. We start by decomposing the convergence field
$\kappa(k,\bl)$ as before. Here $\kappa^O(k,\bl)$ is the observed convergence assuming a fiducial cosmology.
\be
\kappa^O(k,\bl) = \sqrt{2\over\pi} \int {d^2 \rpe \over 2 \pi} \int { dz_p } n_z(z_p)\kappa(\br)kj_l(k r^0) \exp[-i\bl\cdot\theta]  .
\ee

\n
By expressing  $\kappa(\rpa,\rpe)$ in terms of $\kappa(k,\bl)$ we find

\be
\kappa^O(k,\bl) = { 2 \over \pi} \int {d \rpa \over \sqrt{2\pi}} \int dz_p \int {dk' \over \sqrt{2\pi}} kj_l(k\rpa^0) k'j_l(k'\rpa) n_z(z_p)\kappa(k',\bl)
\ee

\be
\kappa^O(k,\bl) = {2 \over \pi} \int dz \int dz_p p(z_p|z) n_z(z_p) k~j_l(k\rpa^0)\int {d\rpa \over {\sqrt {2\pi} }}\int {dk' \over {\sqrt {2\pi} }}k'~j_l(k'r')\kappa(k',\bl).
\ee

\n
This expression enables us to relate the theoretical predictions for a fiducial cosmology
to the observed $\kappa^O(k,\bl)$. We have assumed $n(r)d\rpe d^2\rpa = {1 \over A}\bar n_z(z_p)dz_p d^2\rpa$, where $A$ is the
solid angle of the sky covered. The statistical properties such as the bispectrum and trispectrum of the field $\kappa(k,\bl)$ derived in the main text
can now be used to predict the observed statistics of   $\kappa_0(k,\bl)$. Mixing of modes due to the photometric redshift error will couple
the radial modes, whereas the partial sky coverage mixes angular modes.

\section{Useful Mathematical Relations}

\subsection{Spherical Bessel Functions}

\n
The orthogonality relationship for the spherical Bessel functions is given by the following expression:

\begin{equation}
\int k^2 j_l(kr_1) j_l(kr_2) dk = \left [ {\pi \over 2 (l+1/2)^2} \right ] \delta_{1D}(r_1-r_2).
\label {eq:limber_approx1}
\end{equation}

\n
The extended Limber approximation is also implemented through the following approximate relation \cite{LoAf08}:

\be
\int F(k) j_l(kr_1) j_l(kr_2) dk \sim  \left [ {\pi \over 2 r_1^2} \right ] F \left ( {l \over r_1} \right ) \delta_{1D}(r_1-r_2).
\label{eq:limber_approx2}
\ee

\n
Thus for high $l$ the spherical Bessel functions can be replaced by a Dirac delta function $\delta_{1D}$:

\be
\displaystyle \lim_{x\to\infty} j_l(x) = \sqrt {\pi \over 2l+1} \delta_{1D} \left ( l+ {1 \over 2} - x\right ).
\label{eq:limber_approx3}
\ee
\subsection{Spherical Harmonics}

\n
The completeness relationship for the spherical harmonics is given by:

\be
\sum_{lm} Y_{lm}(\oh)Y_{lm}(\oh') = \delta_{2D}(\oh-\oh').
\label{complete_spherical}
\ee

\n
The orthogonality relationship is as follows:

\be
\int d\oh Y_{lm}(\oh) Y_{l'm'}(\oh) = \delta^K_{ll'}\delta^K_{mm'}.
\label{ortho_spherical}
\ee

\subsection{3J Symbols}

\n
The following properties of $3J$ symbols were used to simplify various expressions.

\be
\sum_{l_3m_3} (2l_3+1) \left ( \begin{array}{ c c c }
     l_1 & l_2 & l_3 \\
     m_1 & m_2 & m_3
  \end{array} \right )
\left ( \begin{array}{ c c c }
     l_1 & l_2 & l \\
     m_1' & m_2' & m
  \end{array} \right ) = \delta^K_{m_1m_1'} \delta^K_{m_2m_2'} \\
\ee

\be
\sum_{m_1m_2} \left ( \begin{array}{ c c c }
     l_1 & l_2 & l_3 \\
     m_1 & m_2 & m_3
  \end{array} \right)
\left ( \begin{array}{ c c c }
     l_1 & l_2 & l_3' \\
     m_1 & m_2 & m_3'
  \end{array} \right) = {\mathcal \delta^K_{l_3l_3'} \delta^K_{m_3m_3'} \over 2l_3 + 1} \\
\ee

\be
(-1)^m\left ( \begin{array}{ c c c }
     l & l & l' \\
     m & -m & 0
  \end{array} \right ) = {(-1)^l \over \sqrt{(2l+1)}} \delta^K_{l'0}.
\label{eq:3j}
\ee

\n

\end{document}